\documentclass[pra,twocolumn,showpacs,superscriptaddress,floatfix]{revtex4-1}
\usepackage{amsmath,amscd,amssymb,color}
\usepackage{graphicx,amsfonts,epsf}
\usepackage{epstopdf}
\usepackage{float}
\usepackage{hyperref}
\usepackage{enumerate}
\usepackage{array}

\usepackage{xcolor}
\begin{document}
\title{State-independent robust
heat-bath algorithmic cooling of nuclear spins}
\author{Krishna Shende}
\email{ph19032@iisermohali.ac.in}
\affiliation{Department of Physical Sciences, Indian
Institute of Science Education \& 
Research Mohali, Sector 81 SAS Nagar, 
Manauli PO 140306 Punjab India.}
\author{Arvind}
\email{arvind@iisermohali.ac.in}
\affiliation{Department of Physical Sciences, Indian
Institute of Science Education \& 
Research Mohali, Sector 81 SAS Nagar, 
Manauli PO 140306 Punjab India.}
\affiliation{Punjabi University Patiala,
147002, Punjab, India}
\author{Kavita Dorai}
\email{kavita@iisermohali.ac.in}
\affiliation{Department of Physical Sciences, Indian
Institute of Science Education \& 
Research Mohali, Sector 81 SAS Nagar, 
Manauli PO 140306 Punjab India.}
\begin{abstract}
In this work, we experimentally demonstrate the
implementation of a recently proposed
robust and state-independent
heat-bath algorithmic cooling (HBAC) method~\cite{Sadegh} on
an NMR quantum processor.  
While HBAC methods improve the purity of a quantum
system via iterative unitary entropy compression, they are
difficult to implement experimentally since they use
sort operations that are different for each iteration.
The new robust HBAC
method proved that optimal HBAC is possible without 
prior state information and using a single fixed
operation. We modified the protocol to experimentally
perform efficient cooling of ${}^{13}$C and ${}^{15}$N spins and
provide an optimal decomposition of this modified protocol
in terms of quantum gates.  This is the first time that
optimal HBAC has been experimentally demonstrated on
${}^{15}$N spins. We examined the relaxation dynamics of
these algorithmically cooled spins, in order to ascertain
the effect 
of decoherence on the cooled states.
\end{abstract} 
\maketitle 
\section{Introduction}
\label{sec1}
Quantum computers have the potential to increase the speed
and efficiency of certain computational algorithms and
simulations~\cite{nielsen,shor,harrow}. However, physical
realizations of quantum computers have proved challenging,
since sensitive quantum mechanical effects are easily
overwhelmed by thermal fluctuations, and small errors during
the implementation of quantum operations are detrimental to
their efficient implementation~\cite{quanteff}.  Protocols
such as quantum error correction and fault tolerant
computation were developed to curb these errors,
which require the supply of pure qubits throughout the
computation process~\cite{qec}.

NMR quantum processors use ensembles of nuclear spins to
perform quantum computational tasks and have the advantages
of long qubit decoherence times and optimal gate
implementation via high-precision rf
pulses~\cite{Cory2000,vandersypen,KD-current-science}.
Nuclear spins at room temperature are in Boltzmann
equilibrium, and thus this highly mixed ensemble  leads to a
major disadvantage for quantum computation which requires
highly pure ensembles~\cite{divincenzo}.  The NMR quantum
information processor has been termed as a system having
`good dynamics versus bad kinematics'~\cite{Linden}.  To
improve this situation, we need to prepare the initial state
of the NMR quantum information processor such that a large
number of spins in the ensemble are in the same quantum
state.  One way to prepare an ensemble of spins in a pure
state is to cool the entire system down to very low
temperatures, which however is not feasible with the current
technology.  Other spin-cooling proposals include
algorithmic cooling~\cite{Schulman}, dynamic nuclear
polarization~\cite{dnp}, parahydrogen induced
polarization~\cite{phip} and optical
pumping~\cite{opticalpump}, all of which have met with
varying degrees of success.

The population difference between the $\vert 0 \rangle$ and
$\vert 1 \rangle$ of an NMR qubit is termed as the `spin
polarization ($\epsilon$)' and at thermal equilibrium:
\begin{equation} \epsilon = P_{\vert 0 \rangle} - P_{\vert 1
\rangle} = {\rm tanh}\left(\frac{\Delta E}{2 k_B T}  \right)
\approx \frac{\hbar \gamma B_z}{2 k_B T} \end{equation}
where $\gamma$ is the gyromagnetic ratio, $B_z$ is the
intensity of the external Zeeman magnetic field, $k_B$ is
the Boltzmann constant, $T$ is the temperature of the bath,
and $\Delta E << k_B T$.  The spin temperature is defined as
$T_{\rm spin} = \frac{\hbar \gamma B_z}{2 k_B \epsilon}$ and
spins with a polarization higher than their room temperature
polarization, can be considered to be `cooled down'.
Cooling of nuclear spins is hence equivalent to increasing
their polarization which is limited by the Shannon
bound~\cite{Cover2006}.

The method for increasing polarization of selected qubits in
a closed system was introduced by Schulman~\cite{Schulman}
and is known as algorithmic cooling (AC).  Here, additional
ancilla qubits are required which are called reset qubits,
and the combined system qubits and reset qubits
are subjected to an overall unitary transformation such that
the system qubits move towards a state with lower
entropy/increased purity and hence with a lower spin
temperature, while the reset qubits heat up and gain
entropy.  The major disadvantage of the AC protocol is that
for spins at room temperature, the number of spins required
to cool the target spin is very large (typically $\approx
10^{12}$ spins).  This obstacle was circumvented by 
Boykin~\cite{Boykin} in an extension to AC called heat-bath
algorithmic cooling (HBAC), wherein a contact between the
heat bath and the system is introduced, which pumps the
excess entropy out of the system into this heat
bath~\cite{jose,schulman3}.  Later an extension to the HBAC
method was proposed, the partner pairing algorithm (PPA),
which sorts the diagonal elements of a density matrix for
entropy compression, and it was proved that this sort
operation is an optimal entropy compression step for
HBAC~\cite{Schulman2}.  If the decoherence time of the reset
qubits is significantly smaller than that of the
computational qubits, it is possible to iteratively achieve
cooling of the target qubits.  However, despite repetitive
cooling, the target qubits cannot be fully purified, and a
limit of purification has been
computed~\cite{elias,asym,pollimit}.  The role of
non-Markovian processes in improving cooling efficiency was
explored  and several protocols were suggested to optimize
the thermalization strategy~\cite{AlhambraHBAC}.  It was
discovered that the unitarity of the compression operation
is what limits the cooling of HBAC
techniques~\cite{Raeisi_No-go}.  It was recently
demonstrated that the cooling limit of HBAC protocols can be
enhanced in the presence of noise~\cite{Farahmand}.  The
overall methodology of all these methods is to perform
various unitary transformations on a multi-spin system where
part of the system moves towards a higher polarization,
lower entropy and hence lower effective spin temperature, at
the cost of the other part which heats up, and in some of
the methods this heat is transported away by another set of
transformations.

On NMR quantum processors, various experiments based on the
HBAC protocol have been
performed~\cite{Baugh,paramagnet,amino,Brassard,yosi,tsm},
some of which showed polarization enhancement beyond
Shannon's limit.  Several iterations of HBAC was performed
on three solid-state NMR qubits in order to cool a single
qubit~\cite{laflamme_multiple}.  The PPA-HBAC algorithm was
used to implement a quantum Otto heat engine
with greater thermal efficiency beyond traditional
engines~\cite{algoQHE}.  An HBAC method using correlated
spin-bath interactions was devised which uses spin-spin
cross-relaxation (the nuclear Overhauser effect) to achieve
higher polarization enhancement as compared to the PPA-HBAC
method~\cite{laflamme-njp}. Recently, HBAC has been used to
enhance spin polarization in NV center quantum
devices~\cite{Zaiser}.

All the previous experimental implementations of HBAC
achieved cooling via the PPA-HBAC method, which requires a
different compression unitary to be implemented as well as
knowledge of the state of the system in every iteration.
The work done in this paper is a realization of Raeisi's
state-independent HBAC technique~\cite{Sadegh} on an NMR
quantum processor, which proposed a new fixed operation to
achieve algorithmic cooling that does not require any prior
knowledge of the state.  We experimentally demonstrate the
successful implementation of this new protocol on two
different three-qubit NMR systems: the first system having a
${}^{13}$C spin as the target qubit to be cooled and the
second system having a ${}^{15}$N spin as the target qubit
to be cooled.  We were able to achieve large polarization
enhancements and concomitantly were able to decrease the
corresponding spin temperatures of the target spins to well
below room temperature.  The original theoretical proposal
decomposed the fixed unitary compression operator in terms
of shift operators, which were further decomposed in terms of
the
quantum Fourier transform (QFT),
the inverse QFT, 
and single-qubit rotation
gates.  
The QFT is a resource-intensive operation whose
circuit depth increases tremendously with the number of
qubits, which makes it more vulnerable to experimental
errors~\cite{kd-ijqi-2005}.  We hence designed an optimal
decomposition of the shift operators in terms of standard
multiqubit gates such as Toffoli, CNOT and NOT gates, which
makes it easier to experimentally implement the circuit.  
After implementing several cycles of TSAC cooling, We
were able to enhance the polarization of the ${}^{13}$C and
${}^{15}$N spins by 4.3 and 5.95 times respectively, as
compared to their thermal equilibrium polarizations, which
translates to cooling their spin temperatures down to
$\approx 71~K$ and $\approx 51~K$, respectively.  It is
noteworthy that ${}^{15}$N spins (due to their low
gyromagnetic ratio) have very low spin polarizations at
thermal equilibrium ($\approx 1/10$th that of ${}^{1}$H
spins), and we were able to considerably enhance their
polarization via the state-independent HBAC protocol. This
is the first experiment that demonstrates cooling of
${}^{15}$N spins using HBAC methods.  Further, we studied
the relaxation dynamics of the algorithmically cooled spins
by measuring their T$_1$ and T$_2$ relaxation rates and
observed that algorithmically cooled states relax in a
way similar to thermal states, thus retaining the good
dynamics of the NMR spin system.

This paper is organized as follows:~The standard HBAC and
PPA-HBAC protocols are described in Section~\ref{sec2a},
while the new state-independent TSAC protocol is given in
Section~\ref{sec2b}.  The optimal circuit decomposition of
the compression unitary is described in Section~\ref{sec2c}.
The experimental implementation of the new state-independent
HBAC protocol is presented in Section~\ref{sec3}.
Section~\ref{sec3a} contains experimental details, while
Sections~\ref{sec3b} and \ref{sec3c} describe the results of
experimentally cooling a ${}^{13}$C-labeled system and a
${}^{13}$C-${}^{15}$N-labeled system, respectively.
Section~\ref{sec4} contains a few concluding remarks. 
\section{Theoretical Framework}
\label{sec2}
\subsection{Standard Heat-Bath Algorithmic Cooling Protocols}
\label{sec2a}
The AC scheme exploits the fact that cooling of specific
spins below their equilibrium polarization can be achieved
by using a reversible unitary operation to increase the
polarization of the target spins, relative to the rest of
the spins.  Closed system AC executes an entropy compression
operation on spins which are initially at thermal
equilibrium  which creates a difference in the spin
temperatures of the spins such that, the reset spins heat up
and the computational spins cool down, as represented
by\cite{Schulman}: 
\begin{equation}
\rho_{\epsilon_i}^{\otimes
n}=\rho\xrightarrow{\text{compression}}\rho^{'}=U\rho
U^{\dagger}=\rho_{\epsilon_c}\otimes\rho_{\epsilon_r}^{\otimes
{n-1}} 
\end{equation} 
where $\epsilon_i$ is the initial
polarization,  and $\epsilon_c$ and $ \epsilon_r$ are the
polarizations of the
computational spin and reset spin respectively, 
$\epsilon_c > \epsilon_i > \epsilon_r$ 
after entropy compression.

The density matrix of the computational spin is obtained by
taking a trace over the $n-1$ reset spin
($\rho_{\epsilon_c}={\rm Tr}_{n-1}(\rho^{\prime})$).  
A major disadvantage of the AC method  is the fact
that cooling is limited by the Shannon bound for information
compression \cite{Cover2006,SORENSEN}.

\begin{figure}[h]
\includegraphics[scale=0.7]{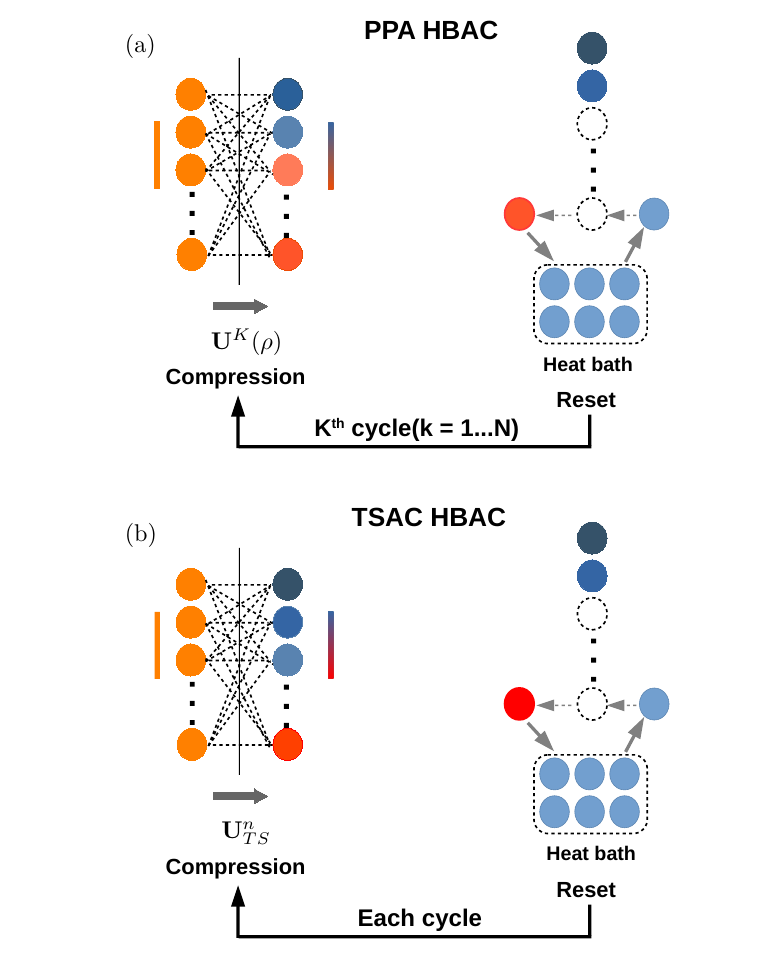}
\caption{Schematic to illustrate the different steps of the
PPA-HBAC and state-independent HBAC protocols.  The $n$ spin
system is categorized into target spins and reset spins.
Temperature is color coded, with shades of blue indicating
cooling and shades of orange indicating heating.  (a) During
every iteration in the PPA HBAC method,  the compression
unitary $U^{1} U^{2}...U^{K}...U^{N}$ extracts entropy from
the target spin and redistributes it among the reset spins,
such that the target spin cools down and the reset spins
heat up.  The compression step is followed by a reset step,
where the hot reset spins are brought into contact with the
heat-bath.  The compression unitary $U^{K}(\rho)$ depends on
the state $\rho$ and changes after every cooling cycle,
where $K=1,2..N$ and $N$ denotes the maximum number of
cooling cycles implemented.  (b) Only a fixed compression
unitary $U^{n}_{{\rm TS}}$ is required in the state-independent TSAC HBAC
method to redistribute entropy and develop a temperature
gradient among spins.  The compression step is followed by a
reset step where the reset spins equilibrate with the
surrounding heat-bath.  The dotted lines in (a) and (b)
indicate redistribution of entropy among the system spins to
create a temperature gradient.}
\label{cartoon}
\end{figure}
\begin{figure*}[ht]
\includegraphics[scale=1.0]{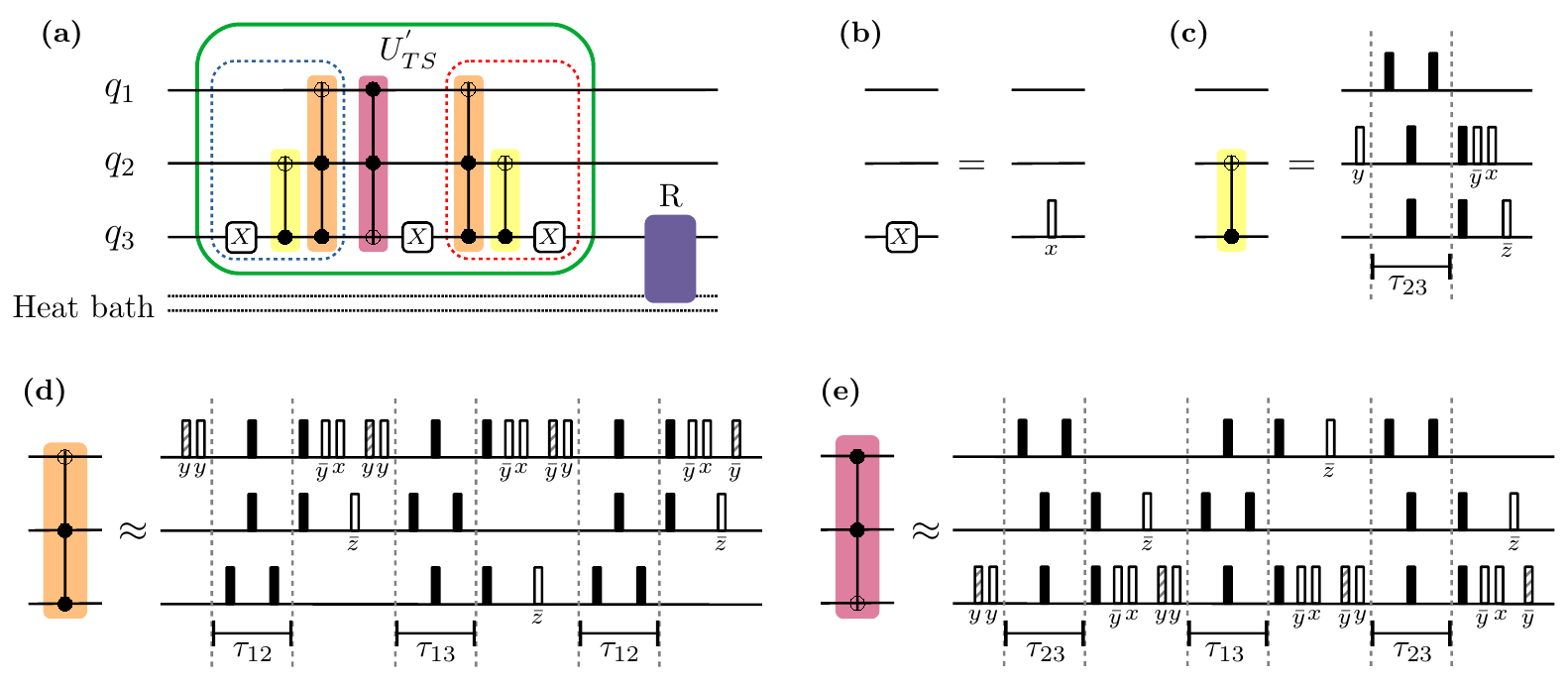}
\caption{(a) Schematic of the quantum circuit to implement a
single cycle of the state-independent HBAC protocol on a
three-qubit system, with $q_i$ labeling the qubit.  The
green box encloses the entire compression unitary operator,
while the dashed blue and red boxes enclose the circuits
corresponding to the right shift and the left shift
operator, respectively; R denotes the reset operation.  The
NMR pulse sequence corresponding to the implementation of a
(b) NOT gate, (c) CNOT gate, (d) Toffoli gate with the
target being the first qubit, and (e) Toffoli gate with the
target being the third qubit, respectively.  The black,
unfilled and cross-hatched rectangles represent $\pi$,
$\frac{\pi}{2}$, and $\frac{\pi}{4}$ spin-selective RF
pulses, respectively.  The phase of the pulse is written
below the respective pulse, with a bar over a phase denoting
negative phase; the delays are denoted by
$\tau_{ij}=\frac{1}{2 {\rm J}_{ij}}$, where $i,j$ are the
qubit labels and ${\rm J}_{ij}$ is the strength of the
scalar coupling.} 
\label{circuit}
\end{figure*}
Cooling in the AC scheme can be enhanced by incorporating a
contact between a heat bath and the system.  The excessive
heat in the reset spins is then pumped out of the system
into a heat bath, where it is removed and the spins are
cooled to the bath temperature a step  called the 'reset'
step \cite{Boykin}.  The system spins are categorized as a
target spin  (which is to be cooled), a set of scratch spins
(which can either be a higher-dimensional qudit or a string
of qubits) which help in entropy compression, and a set of
$m$ reset spins which are brought into contact with the
heat-bath~\cite{pollimit}.  The target and scratch spins
together are referred to as the computational spins.  The
entire process of entropy compression and the reset step is
known as heat bath algorithmic cooling
(HBAC)\cite{jose}\cite{schulman3}. 
The HBAC method works optimally if the ratio $R$ between the
thermalization times (T$_1$) of computational to reset spins
satisfies $R >> 1$, so that several cooling cycles can be
implemented, and the state of the computational spins
remains unchanged during the reset process.  The reset step
amounts to tracing over the reset spins and replacing them
with heat-bath spins.  It is assumed that the heat-bath
capacity is so large that the spin-bath interaction has no
effect on the bath temperature.

The HBAC method was further optimized in a method called the
partner pairing(PPA) algorithm~\cite{Schulman2}, where in
each iteration the diagonal elements of the density matrix
are sorted in decreasing order.  It is to be noted that the
sort operations and their complexity keep changing from
iteration to iteration.  

The steps of the PPA HBAC algorithm can be written as:
\begin{eqnarray}
\rho  \stackrel{{\rm C}}{\longrightarrow} \rho^{\prime}
&=& U_C \rho U_C^{\dagger}  \nonumber \\
\rho^{\prime}  \stackrel{{\rm S}}{\longrightarrow}
\rho^{\prime\prime}
&=& U_S(\rho^\prime) \rho^{\prime}
U_S^{\dagger}(\rho^{\prime})  \nonumber \\
\rho^{\prime\prime} \stackrel{{\rm R}}{\longrightarrow}
\rho^{\prime \prime\prime}
&=&
{\rm Tr}_{m}\left[ {\mathcal L}(\rho^{\prime\prime} \otimes 
\rho_{{\epsilon_{b}}}^{\otimes m})\right]
\label{ppa-steps}
\end{eqnarray}
Here $U_C$ and $U_S$ are  unitaries involved in compression and
sorting steps and act on the system qubits, while ${\mathcal
L}$
represents non-unitary evolution during the period  in which
the reset qubits cool down due to interaction with the 
heat bath;
$\rho_{\epsilon_{b}}$ represents
the state of the bath spin and $\epsilon_b$ is the
heat-bath polarization.
The cooling limit is achieved once no more entropy extraction is
possible i.e. when system state reaches a steady state which is not
changed by compression and refresh 
($\rho = \rho^{\prime \prime \prime}$ in Eq.~(\ref{ppa-steps}))
For $n$ spins the PPA HBAC cooling limit
i.e. the maximum achievable polarization is computed as~\cite{laflamme-njp}:
\begin{equation}
\epsilon_{{\rm max}} = 
\frac{(1+\epsilon_b)^{2^{n-2}} - (1 - \epsilon_b)^{2^{n-2}}}
{(1+\epsilon_b)^{2^{n-2}} + (1 - \epsilon_b)^{2^{n-2}}}
\label{max-cool}
\end{equation}
Spin temperature is inversely proportional to the
polarization and can be computed from:
\begin{equation}
{\rm T}_1 . \epsilon_1 = {\rm T}_2 . \epsilon_2
\label{spin-temp-eqn}
\end{equation}
where ${\rm T}_1, {\rm T}_2$ are the initial and final spin
temperatures and
$\epsilon_1, \epsilon_2$ are the initial and final spin
polarizations, respectively.

An upper bound on spin cooling (i.e. on polarization enhancement),
also called the \emph{Shannon bound},
can be derived
by interpreting the spin state in terms of information theory, wherein the
information content ($IC$) of the spin is defined using Shannon entropy $H$,
and the relation between spin polarization ($\epsilon$) and $IC$ is given
by~\cite{pollimit}:
\begin{equation}
\begin{split}
H_{1qubit}=&\biggr[\frac{1-\epsilon}{2}\ln\left(\frac{1-\epsilon}{2}\right) +\frac{1+\epsilon}  {2}\ln\left(\frac{1+\epsilon}{2}\right)\biggr]\\
IC_{1qubit}=&1-H_{1qubit}=\frac{\epsilon^{2}}{\ln4}+O(\epsilon^4).
\end{split}
\label{shannon-eqn}
\end{equation}

A major limitation of the PPA algorithm is that in each
iteration, complete information about the state of the
system is needed in order to set up the sort operation. In a
domino effect, this in turn implies that the PPA compression
operation also changes after every iteration, which makes
PPA HBAC method experimentally challenging to implement.

A schematic of the PPA HBAC method is given in
Fig.~\ref{cartoon}(a) wherein an entropy compression
operation is first implemented which arranges the diagonal
elements of the density matrix in decreasing order.  These
sort operations are state dependent and hence the unitaries
used in each compression step of the PPA HBAC method are
different.  The excess heat of the reset spins is pumped out
of the system by making them interact with the heat bath. 

\subsection{Two-Sort Algorithmic Cooling}
\label{sec2b}
A new HBAC method was recently proposed  (termed the
two-sort algorithmic cooling (TSAC) method) which achieves
optimal cooling of the system spins, and has been shown to
be better than the PPA HBAC method in terms of circuit
complexity and robustness against noise~\cite{Sadegh}.
Cooling is achieved by recursively operating a novel fixed
unitary matrix followed by the reset step, without requiring
prior state information.  The general unitary matrix for $n$
spins ($U_{{\rm TS}}$) is given by:
\begin{equation}
U^{n}_{{\rm TS}}=\begin{bmatrix}
1 &  &  &  &  &  &   \\
& X &  & &  &  &    \\
&  & . &  &  &  &    \\
&  &  & . &  &  &    \\
&  &  &  & . &  &    \\
&  &  &  &  & X &    \\
&  &  &  &  &  & 1  \\
\end{bmatrix}
\label{nmatrix}
\end{equation}
where $X$ denotes the Pauli $\sigma_{x}$ matrix.  Unlike the
unitary compression operator in the PPA algorithm which
achieves a descending sort of the diagonal elements, the
two-sort unitary $U_{TS}$ swaps every two neighboring
diagonal elements, except for the first and last elements.
The matrix is $2^{n} \times 2^{n}$ and acts on the
computation and reset spins, achieving a local partial sort
of the density matrix diagonal elements.

For three spins, the unitary compression operator is given by:
\begin{equation}	
U^{3}_{{\rm TS}}=\begin{bmatrix}
1 & 0 & 0 & 0 & 0 & 0 & 0 & 0 \\
0 & 0 & 1 & 0 & 0 & 0 & 0 & 0 \\
0 & 1 & 0 & 0 & 0 & 0 & 0 & 0 \\
0 & 0 & 0 & 0 & 1 & 0 & 0 & 0 \\
0 & 0 & 0 & 1 & 0 & 0 & 0 & 0 \\
0 & 0 & 0 & 0 & 0 & 0 & 1 & 0 \\
0 & 0 & 0 & 0 & 0 & 1 & 0 & 0 \\
0 & 0 & 0 & 0 & 0 & 0 & 0 & 1 \\
\end{bmatrix} 
\label{3q_matrix}
\end{equation} 

The polarization of spins depends on the values of the
diagonal elements of the density matrix of the system.  The
unitary matrix $U_{{\rm TS}}$ (Eq.~(\ref{nmatrix})) keeps
the first and the last diagonal element of the density
matrix unchanged, and exchanges the position of the other
elements with their immediate neighbors.  By choosing a spin
with low polarization as the first spin, and performing
$U_{{\rm TS}}$ iteratively on a system in thermal
equilibrium, the polarization of the first spin can be
substantially enhanced. 

A schematic diagram of the TSAC HBAC method is given in
Fig.~\ref{cartoon}(b), where a single unitary $U^{n}_{{\rm TS}}$ is used to
compress entropy among spins, thereby creating a temperature
gradient between the system spins. The target spin is cooled
while the reset spins are heated up.  The reset step  is the
same for the PPA HBAC and the TSAC HBAC protocols.  The reset
step is repeated several times to thermalize the reset spins
to the heat-bath temperature. 
\subsection{Optimal Circuit Decomposition of the Compression Unitary}
\label{sec2c}
The first and last operations of the $U_{TS}$ unitary
operator (the $1 \times 1$ blocks in the top left and bottom
right corners of the matrix) in Eq.~(\ref{nmatrix})
correspond to SHIFT$_m$ operators which shift the spin state
$m$ times to the right or the left.  This can be achieved by
applying a multiple-control-Toffoli gate (which is a
controlled-controlled-NOT with one target and $n-1$
controls) followed by a NOT gate on the last spin.  The
suggested decomposition in Raeisi et.~al.~\cite{Sadegh} for
these SHIFT operators is a QFT and an inverse QFT,
sandwiching a set of rotation operators of specific rotation
angles and phases.

In order to use fewer experimental resources and alleviate
the detrimental effects of noise during long gate
implementation times, we have decomposed the SHIFT operator
into a sequence of a three-qubit Toffoli gate, a two-qubit
CNOT gate and a single-qubit NOT gate.  The complexity of
the circuit has been further reduced by using an approximate
Toffoli gate as described in Ref.~\cite{gate}, which differs
from the actual Toffoli gate by a phase of one of its
amplitudes (the phase of the $|101\rangle$ state is
reversed).  The final compression unitary operator which has
been experimentally implemented is given by:
\begin{equation}        
U^{\prime}_{{\rm TS}}=\begin{bmatrix}
1 & 0 & 0 & 0 & 0 & 0 & 0 & 0 \\
0 & 0 & 1 & 0 & 0 & 0 & 0 & 0 \\
0 & 1 & 0 & 0 & 0 & 0 & 0 & 0 \\
0 & 0 & 0 & 0 & 1 & 0 & 0 & 0 \\
0 & 0 & 0 & 1 & 0 & 0 & 0 & 0 \\
0 & 0 & 0 & 0 & 0 & 0 & -1 & 0 \\
0 & 0 & 0 & 0 & 0 & 1 & 0 & 0 \\
0 & 0 & 0 & 0 & 0 & 0 & 0 & 1 \\
\end{bmatrix}  
\label{3q_mat}
\end{equation}
The negative sign at the $|101\rangle\langle110|$ position
in the $U^{\prime}_{{\rm TS}}$ unitary operator (Eq.~(\ref{3q_mat})) has no
effect on the TSAC protocol and the results remain
the same. 

The complete quantum circuit for implementing the TSAC
protocol is depicted in Fig.~\ref{circuit}(a),  for a
three-qubit system initially prepared in a thermal
equilibrium state.  The circuit inside the green box is the
general compression unitary $U^{3}_{{\rm TS}}$, as given in
Eq.~(\ref{3q_mat}).  This unitary is decomposed as a
sequence of a right SHIFT operator, a left SHIFT operator, a
Toffoli gate and a NOT gate, enclosed in the blue dashed
box, the red dashed box, the magenta shaded and the unfilled
box, respectively, in Fig.~\ref{circuit}(a). Each SHIFT
operator is further decomposed as a Toffoli gate, a CNOT
gate and a NOT gate given in the orange shaded, yellow
shaded and unfilled boxes, respectively.  The reset step of
the TSAC algorithm is denoted by the indigo shaded box,
with the horizontal lines depicting the heat bath.
Figs.~\ref{circuit}(b)-(e) depict the NMR pulse sequences
for a NOT, a CNOT and a Toffoli gate, respectively.
\section{Experimental implementation}
\label{sec3}
\subsection{Experimental details}
\label{sec3a}
All experiments  were performed  at  ambient  temperature (303~K) on a Bruker
Avance III 600-MHz NMR spectrometer equipped with a standard 5 mm TXI probe. 
The Hamiltonian for a three-spin system in the rotating frame, assuming
a high-temperature and high-field approximation, is given
by~\cite{OLIVEIRA}:
\begin{equation}
\mathcal{H}=-\hbar\sum\limits_{i=1}^3 { \omega _i } I_z^i +
\hbar\sum\limits_{i<j=1}^3  J_{ij} I_z^i I_z^j
\end{equation}
where $\omega_i, I_z^i$ represent the offset frequency and the $z$-component of
the spin angular momentum of the $i$th spin respectively, and $J_{ij}$ is the
strength of the scalar coupling between the $i$th and $j$th spins.

The TSAC cooling protocol was experimentally implemented on two molecules.  
The first molecule was ${}^{13}$C$_2$-labeled glycine with the
spin denoted as ${}^{13}$C1 being the target spin to be cooled 
and the $^{1}$H spin being the reset spin (see
Fig.~\ref{13C-spectrum}(a) for the molecular structure, chemical shifts $\nu_i$
and $J_{ij}$ scalar coupling values).  
The glycine molecule was dissolved in D$_2$O and a paramagnetic
reagent Cr(acac)$_3$ was added to improve the T$_1$ ratio between the target
and reset nuclei.
This molecule is an example of an A$_2$XX$^{\prime}$ spin system, with
two equivalent ${}^{1}$H spins (A$_2$) and two 
magnetically inequivalent ${}^{13}$C spins (XX$^{\prime}$). 
The  second molecule was
$^{13}$C-$^{15}$N-labeled formamide, with both the $^{15}$N and the ${}^{13}$C
spins being the targets to be cooled, and the $^{1}$H spin being the reset spin
(see Fig.~\ref{15N-spectrum}(a) for the molecular structure, chemical shifts
$\nu_i$ and $J_{ij}$ scalar coupling values).  Each single-qubit rotation gate
was implemented using spin-selective rf pulses  of appropriate phase, power and
time duration, while two-qubit and three-qubit gates were implemented via
evolution under the system Hamiltonian using time delays interspersed with $\pi$
pulses to refocus chemical shifts and retain only the desired scalar coupling
interactions.  More experimental details of NMR pulse sequences for various
quantum gates used in this work can be found in
References~\cite{kd-pra-2015-2,kd-pra-2018-1,kd-pra-2018-3}.  
On the TXI probe,
the duration of
the $\frac{\Pi}{2}$ pulses for $^{15}$N, 
${}^{13}$C and $^{1}$H were 38~$\mu$s at
a power level of 246.6~W, 12.95~$\mu$s  at a power level of  237.3~W, and
7.3~$\mu$s at a power level of 19.9~W, respectively.  
Gradient ascent pulse engineering (GRAPE), an optimal control algorithm, 
was used to generate high-fidelity rf pulses of duration $\approx 150\mu$~s 
to implement single-qubit rotations on the ${}^{13}$C$_2$-labeled glycine
system.
The total time taken to
implement the compression unitary on the $^{13}$C$_2$-labeled glycine
system  and on the $^{13}$C-$^{15}$N-labeled formamide system was 0.302 s and
0.23 s respectively,  which is much shorter than the relaxation times of all the
spins in both systems.
\subsection{Experimentally Cooling the ${}^{13}$C Spin}
\label{sec3b}
\begin{figure}[ht]
\includegraphics[scale=1]{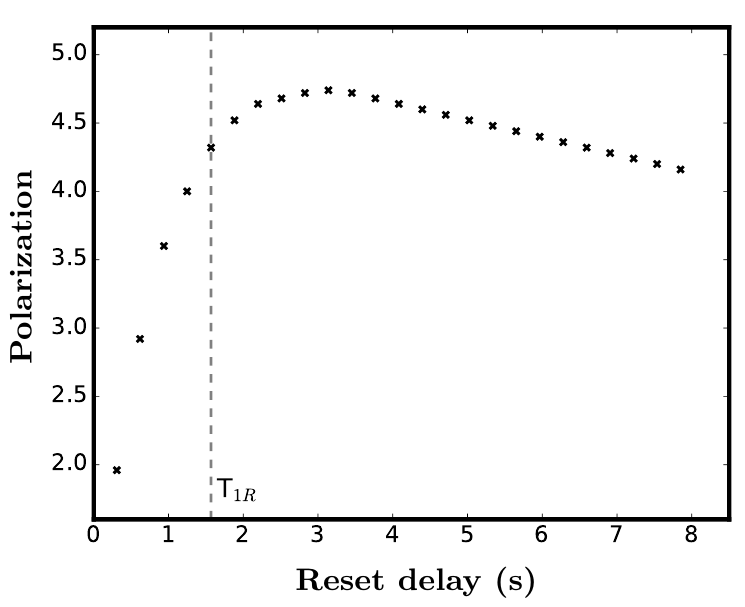}
\caption{Simulation of ${}^{13}$C$1$ spin polarization as a function of the TSAC
reset delay for the ${}^{13}$C$_2$-labeled glycine system; T$_{1R}$ marks
the T$_1$ relaxation time of the ${}^{1}$H reset spin.
}
\label{13Cdelay}
\end{figure}

Most previous experimental algorithmic cooling protocols have focused on cooling
${}^{13}$C spins, due to their small gyromagnetic ratios and correspondingly low
initial polarization at thermal equilibrium ($\approx 1/4$th that of the
${}^{1}$H spin).  We hence 
chose ${}^{13}$C$1$ as the target spin to be cooled in the
$^{13}$C$_2$-labeled glycine system and  the ${}^{1}$H spin as the reset
spin due to its high polarization (set to 1.0) at thermal equilibrium and its
fast T$_1$ relaxation time as compared to the ${}^{13}$C1 spin.  Interestingly,
this system has another ${}^{13}$C2 spin, 
whose polarization also has the potential
to be enhanced.  However, as will be seen below, due to its unfavorable
relaxation properties this spin does not cool down to the same extent as the
target spin.

\noindent{\bf Optimizing the TSAC reset delay:}\\
We simulated the change in ${}^{13}$C1 spin polarization as a function of the
reset delay in order to find the optimum reset delay for iterative TSAC cooling.
We assumed ideal
compression gate implementation and accounted for the decay in the
polarization during the reset delay which is governed by:
\begin{equation}
\epsilon_{t}=(\epsilon_{{\rm init}}-\epsilon_{{\rm eq}})
\exp^{-t/T_1}+\epsilon_{{\rm eq}} 
\label{reset-equation}
\end{equation}
where $\epsilon_{{\rm init}}$ and $\epsilon_{{\rm eq}}$ are the initial and
thermal equilibrium spin polarizations, respectively.  The reset delay was
varied from $0.2*$T$_{1R}$ to $5*$T$_{1R}$, where T$_{1R}$ is the longitudinal
relaxation time of the reset qubit, and as can be seen from Fig.~\ref{13Cdelay},
maximum ${}^{13}$C1 polarization was obtained for a reset delay of 3.14 s, which
is $\approx 2 T_{1R}$. We hence set the reset delay time during heat-bath
interaction to this optimal value in our experiments and performed multiple
rounds of TSAC cooling.
\begin{figure}[ht]
\includegraphics[scale=1]{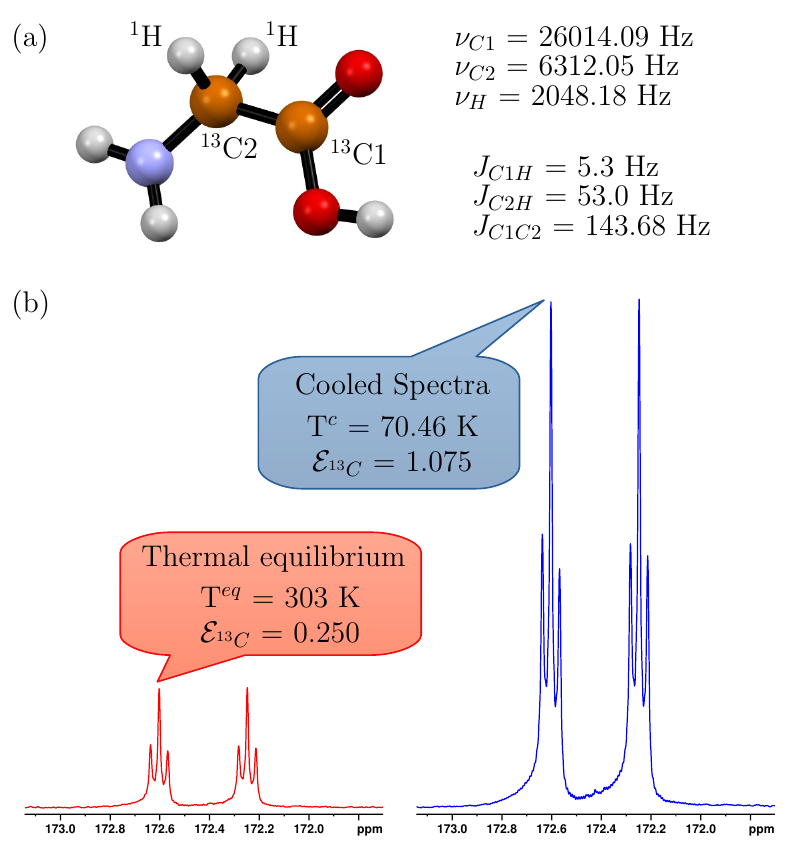}
\caption{(a) Molecular structure of $^{13}$C$_2$-labeled glycine 
with the $^{13}$C1, $^{13}$C2 
and $^{1}$H spins, encoded as the first, second and
third qubit,  respectively. The offset rotation frequency for each spin and
scalar coupling strengths are listed alongside.   (b) The ${}^{13}$C  spectrum
(red) on the left was recorded at thermal equilibrium while the spectrum (blue)
on the right is the enhanced polarization spectrum recorded after algorithmic
cooling  of both the $^{13}$C spins.  The corresponding 
polarizations and spin temperatures are
annotated.  }
\label{13C-spectrum}
\end{figure}

\begin{figure}[ht]
\includegraphics[scale=1]{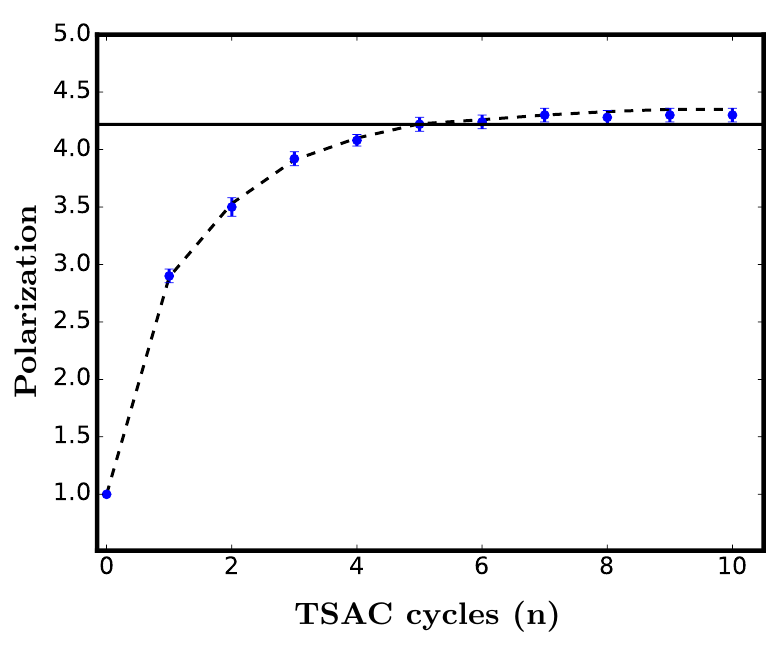}
\caption{Polarization of ${}^{13}$C1 spin 
versus number of cycles (${\bf n}$) of
TSAC algorithmic cooling implemented on the ${}^{13}$C$_2$-labeled glycine
system. Theoretically expected and experimentally obtained
values are depicted by a black dashed line and blue dots, respectively.
The solid black line denotes the {\em Shannon bound} for maximum
achievable polarization.
}
\label{13Ccooling}
\end{figure}

\noindent{\bf Spin temperature after several TSAC cycles:}\\
The ${}^{13}$C spectra at thermal equilibrium and after implementation of
ten rounds of 
the TSAC cooling procedure are shown in Fig.~\ref{13C-spectrum}(b), with  `hot'
thermal equilibrium spectra shown in red and the algorithmically cooled spectra
in blue, plotted to the same scale.  The ${}^{13}$C spectra obtained after
algorithmic cooling using the TSAC method show a considerable increase in spin
polarization ($\approx 4.3$ times), with peak intensities being much higher
than those obtained at thermal equilibrium.  The population of an energy level
(and hence the polarization bias) is directly proportional to the area under the
corresponding resonance peak and was computed via integration.  The theoretical
value of maximum achievable ${}^{13}$C polarization ({\em Shannon bound}) 
for the
${}^{13}$C$_2$-labeled glycine system
computed from Eq.~(\ref{shannon-eqn}) is given by:
\begin{equation}
\begin{split}
IC_{{\rm eq}}=17.84&\frac{\epsilon_{{\rm eq}}^2}{\ln4}
=\frac{\epsilon_{{\rm max}}^2}{\ln4}\\
\Rightarrow \epsilon_{{\rm max}}&=4.224 \epsilon_{{\rm eq}}
\end{split}
\label{13c-maxcooling}
\end{equation}
After ten rounds of TSAC cooling, we were able to experimentally achieve a final
polarization of $\approx 4.3$ for the ${}^{13}$C1 spin, and were clearly able
to surpass the {\em Shannon bound} for this system.  The final spin temperatures
attained by the target  ${}^{13}$C1 spin, the second ${}^{13}$C2 spin and the
reset spin in the ${}^{13}$C$_2$-labeled glycine system are tabulated in
Table~\ref{13c-spin-temp}.  It can be seen that the target spin has been
substantially cooled down to $\approx 71$~K.
The buildup of ${}^{13}$C1 spin polarization after implementation of every cycle
of TSAC cooling is shown in Fig.~\ref{13Ccooling}, and saturation of the
polarization enhancement is attained after four cycles of TSAC cooling.
\begin{table}[h]
\begin{tabular}{|c|c|c|c|}
\hline
Spin & Initial Polarization ($\epsilon_1$)  
& Final Polarization ($\epsilon_2$) & T$(K)$  \\ 
\hline
C1 & 0.25 & 1.075 & 70.46 \\ 
\hline
C2 & 0.25 & 0.687 & 110.26 \\ 
\hline
H & 1 & 0.248 & 1221.77 \\ 
\hline
\end{tabular}
\caption{Initial and final spin polarizations ($\epsilon_1,\epsilon_2$) 
and final spin temperature (T($K$))
attained by each spin in the 
three-spin $^{13}$C$_2$-labeled glycine system, after ten
rounds of algorithmic cooling.}
\label{13c-spin-temp}
\end{table}

\noindent{\bf Magnetization trajectories during compression unitary 
implementation:}\\
The plots of individual spin magnetizations of the $^{13}$C$_2$-labeled glycine
system at the end of implementation of each gate in the quantum circuit of
the TSAC algorithm are shown in Fig.~\ref{mag-plot}(a).  The way entropy
compression proceeds after each gate can be visualized from these magnetization
plots, with the magnetization being plotted on the $y$-axis, and the gate number
being denoted along the $x$-axis; gate number 1, 2, 3, 4, 5, 6, 7 and 8
corresponds to a NOT, CNOT, Toffoli, Toffoli, NOT, Toffoli, CNOT and NOT gate,
respectively (Fig.~\ref{circuit}(a)).  The first three gates comprise the right
shift operator, after implementation of which, the entropy of the ${}^{13}$C1
spin is decreased, that of the ${}^{13}$C2 spin is maximized  and that of the
${}^{1}$H spin remains the same.  As seen from Fig.~\ref{mag-plot}(a), the sixth
gate which is a Toffoli gate has achieved entropy compression and complete
increase of  ${}^{13}$C1 polarization. The subsequent gates manipulate the
${}^{13}$C2 magnetization and invert the ${}^{1}$H magnetization; the
protocol could have been truncated at the seventh gate, which would have reduced
the circuit complexity.
\subsection{Experimentally Cooling the ${}^{15}$N Spin}
\label{sec3c}
\begin{figure}[ht]
\includegraphics[scale=1]{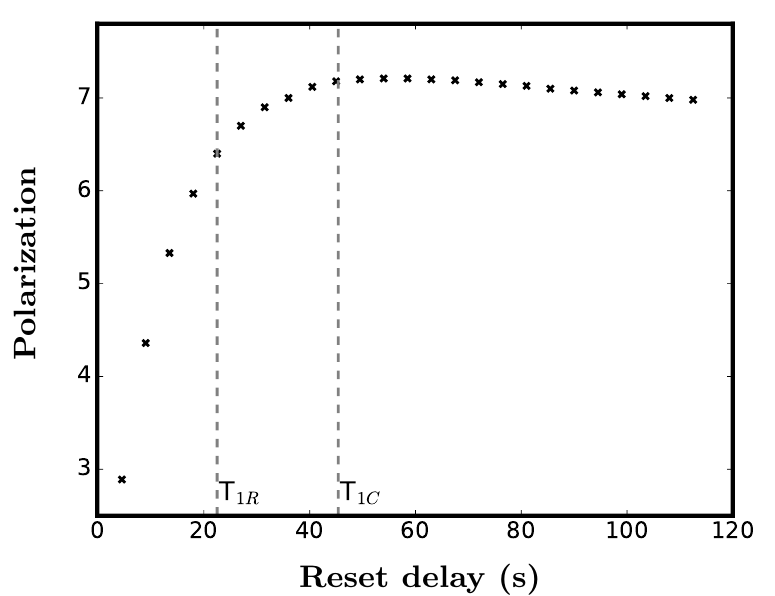}
\caption{Simulation of ${}^{15}$N spin polarization as a function of the TSAC
reset delay for the ${}^{13}$C-${}^{15}$N-labeled formamide system; 
T$_{1R}$  and T$_{1C}$  mark the T$_1$ relaxation times of the ${}^{1}$H
reset spin and the ${}^{15}$N computation spin, respectively.
}
\label{15Ndelay}
\end{figure}
We were interested in exploring the efficiency of the TSAC algorithm on
${}^{15}$N spins, due to their low initial polarizations at thermal equilibrium
($\approx 1/10$th that of the ${}^{1}$H spin) as well as long T$_1$ relaxation
rates, which makes them attractive target spins for cooling. 
Hence ${}^{15}$N was chosen as the main target spin to be cooled in the
${}^{13}$C-${}^{15}$N-labeled formamide system and ${}^{1}$H was chosen as the
reset spin.  This system also contains a ${}^{13}$C spin which can
also potentially be cooled down by the TSAC protocol. 
The reset spin relaxes much faster 
as compared to the  other spins 
and hence can be made to equilibrate quickly with the
heat-bath while the target spin remain relatively isolated from the heat-bath.
 
\noindent{\bf Optimizing the TSAC reset delay:}\\
In order to optimize the reset delay at the end of the compression unitary
implementation in the TSAC protocol, we simulated the change in ${}^{15}$N spin
polarization  as a function of the reset delay.  The TSAC reset delay was varied
from $0.2*$T$_1$ to $5*$T$_1$ for the reset spin (Eq.~(\ref{reset-equation})) and as can
be seen from Fig.~\ref{15Ndelay}, maximum ${}^{15}$N polarization was obtained
for a reset delay of $\approx 2.5$T$_1$. We hence set the value of the reset
delay to this optimal value and performed multiple rounds of TSAC cooling.

\begin{figure}[ht]
\includegraphics[scale=1]{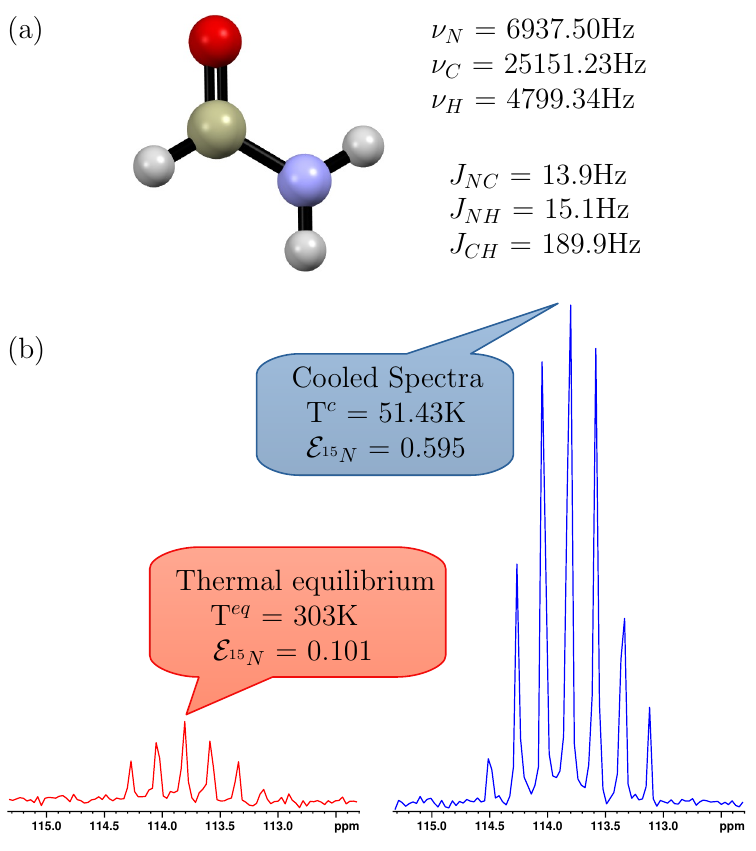}
\caption{(a) Molecular structure of $^{13}$C-$^{15}$N-labeled formamide, with
the $^{15}$N, $^{13}$C and $^{1}$H spins encoded as the first, second and third
qubit, respectively.  The offset rotation frequency for each spin and the scalar
J coupling strengths are listed alongside.   (b) The ${}^{15}$N NMR spectrum
(red) on the left was recorded at thermal equilibrium while the spectrum (blue)
on the right is the enhanced polarization spectrum recorded after algorithmic
cooling  of the $^{15}$N spin.  The corresponding 
polarizations and spin temperatures are
annotated.  }
\label{15N-spectrum}
\end{figure}

\begin{figure}[ht]
\includegraphics[scale=1]{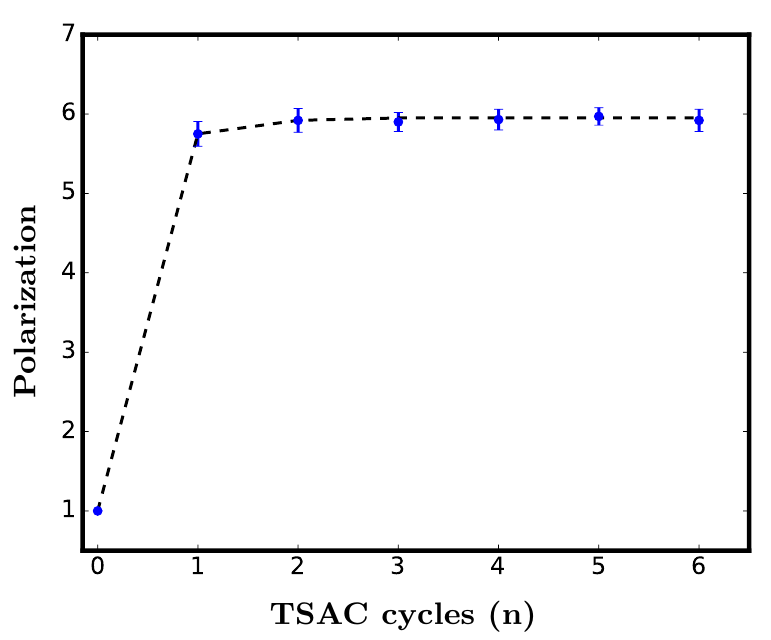}
\caption{Polarization of the ${}^{15}$N spin versus number of 
cycles (${\bf n}$) of
TSAC algorithmic cooling 
implemented on the ${}^{13}$C-${}^{15}$N-labeled formamide
system. Theoretically expected and experimentally obtained values are
depicted by a black dashed line and blue dots, respectively.
}
\label{15Ncooling}
\end{figure}

\noindent{\bf Spin temperature after several TSAC cycles:}\\
The ${}^{15}$N spectra at thermal equilibrium (spectra plotted in red color) and
after implementation of the TSAC cooling procedure (spectra plotted in blue
color) are shown in Fig.~\ref{15N-spectrum}(b).  A substantial increase in
polarization of the ${}^{15}$N was achieved after TSAC cooling ($\approx 5.95$
times). As seen from the spectra in Fig.~\ref{15N-spectrum}(b),  the spectral
peaks at the extreme positions in the thermal equilibrium spectra  are barely
visible, while after algorithmic cooling their signal-to-noise ratio has been
significantly increased.  
The theoretical value of maximum achievable ${}^{15}$N
polarization ({\em Shannon bound}) for the ${}^{13}$C-${}^{15}$N-labeled formamide system
computed from Eq.~(\ref{shannon-eqn}) is given by:
\begin{equation}
\begin{split}
IC_{{\rm eq}}=104.56&\frac{\epsilon_{{\rm eq}}^2}{\ln4}
=\frac{\epsilon_{{\rm max}}^2}{\ln4}\\
\Rightarrow \epsilon_{{\rm max}}&=10.22\epsilon_{{\rm eq}}
\end{split}
\label{15Nmaxcooling}
\end{equation}
After six rounds of TSAC cooling, 
we were able to
experimentally achieve a final ${}^{15}$N polarization of 0.595, 
and were not able to surpass the {\em Shannon bound} for this
system.
The final spin temperatures attained by all the three spins
in the ${}^{13}$C-${}^{15}$N-labeled formamide system are tabulated in
Table~\ref{15n-spin-temp}.  While the ${}^{15}$N spin shows substantial cooling
down to $\approx 51$~K, it is noteworthy that 
the ${}^{13}$C spin has also been cooled 
down to $\approx 187$~K. 
The buildup of ${}^{15}$N spin polarization after implementation of every cycle
of TSAC cooling is shown in Fig.~\ref{15Ncooling}, and saturation of the
polarization enhancement is attained after three cycles of TSAC cooling.
\begin{table}[h]
\begin{tabular}{|c|c|c|c|}
\hline
Spin & Initial Polarization ($\epsilon_1$)  
& Final Polarization ($\epsilon_1$) & T$(K)$  \\ 
\hline
N & 0.101 & 0.595 & 51.43 \\ 
\hline
C & 0.251 & 0.406 & 187.32 \\ 
\hline
H & 1 & -0.187 & $\vert 1620.32 \vert$ \\ 
\hline
\end{tabular}
\caption{Initial and final spin polarizations ($\epsilon_1,\epsilon_2$) 
and final spin temperature (T($K$))
attained by each spin in the 
three-spin  
${}^{13}$C-${}^{15}$N-labeled formamide system, after several
rounds of algorithmic cooling.}
\label{15n-spin-temp}
\end{table}
\begin{figure}[ht]
\includegraphics[scale=1.0]{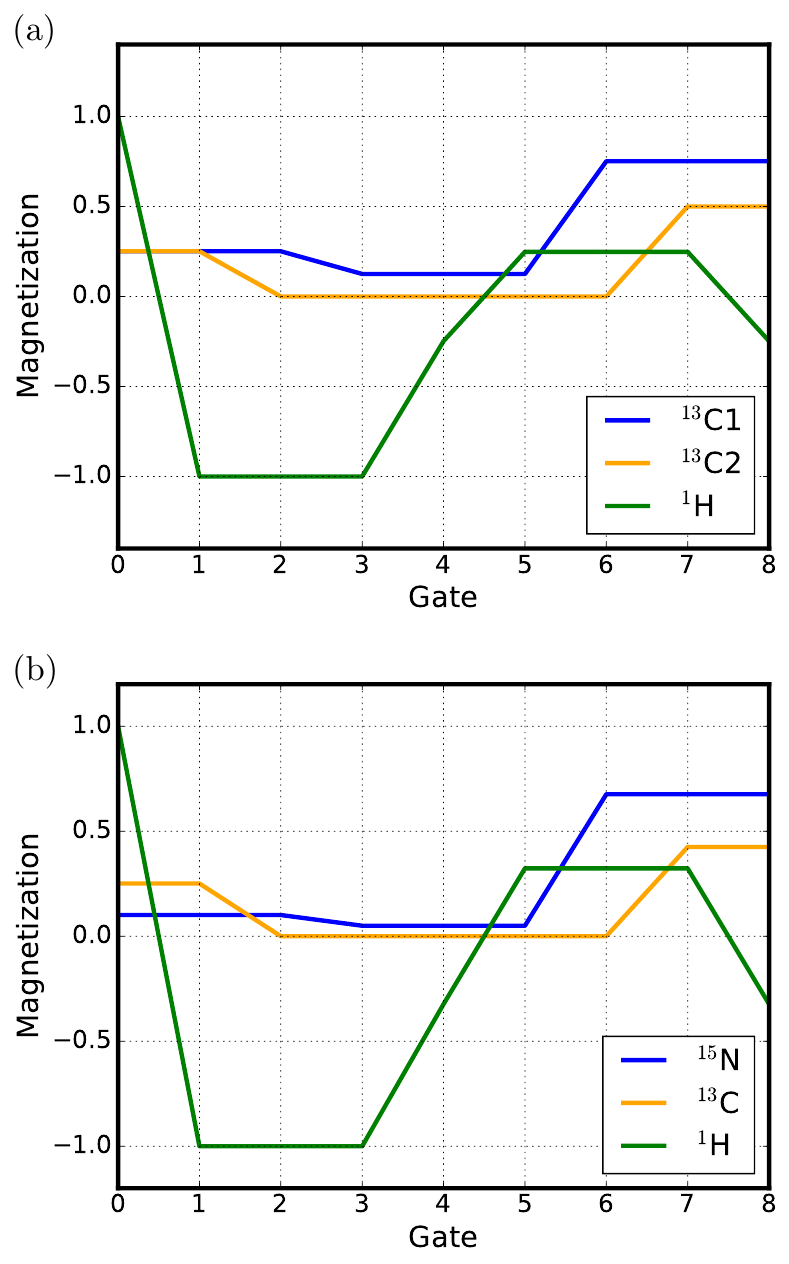}
\caption{Plots of individual spin magnetizations after implementation of
each quantum gate in the circuit of the compression unitary
$U_{TS}$ for the
(a) $^{13}$C$_2$-labeled glycine system
and the
(b) $^{13}$C-$^{15}$N-labeled formamide system.
The gate number is indicated on the $x$-axis.}
\label{mag-plot}
\end{figure}
\begin{table}[h]
\begin{tabular}{ |c|c | c|c |c| }
\hline
&    H &  C2  & C1$^{{\rm hot}}$ & C1$^{{\rm cold}}$  \\ 
\hline
T$_1$  & 1.57$\pm$0.01 &  3.23$\pm$0.03  & 20.4$\pm$0.57 & 18.6$\pm$0.4\\ 
\hline
T$_2$  & 1.0$\pm$0.01  & 1.16$\pm$0.02 &  1.53$\pm$0.01 & 1.4$\pm$0.01\\ 
\hline
\end{tabular}
\caption{Measured relaxation times (in seconds)  of all the spins in the
three-spin $^{13}$C$_2$-labeled glycine system, including the target
spin at thermal equilibrium (C1$^{{\rm hot}}$) and after algorithmic cooling
(C1$^{{\rm cold}}$).}
\label{13c-relax-table}
\end{table}
\begin{table}[h]
\begin{tabular}{ |c|c | c|c |c| }
\hline
&    H &  C  & N$^{{\rm hot}}$ & N$^{{\rm cold}}$ \\
\hline
T$_1$  & 22.5$\pm$0.675 &  30.40$\pm$1.55  & 45.35$\pm$2.25 & 50.67$\pm$3.39\\
\hline
T$_2$  & 1.15$\pm$0.16  & 1.33$\pm$0.07 &  0.095$\pm$0.009 & 0.115$\pm$0.01 \\
\hline
\end{tabular}
\caption{Measured relaxation times (in seconds)  of all the spins in the
three-spin $^{13}$C-${}^{15}$N-labeled formamide system, including the target
spin at thermal equilibrium (N$^{{\rm hot}}$) and after algorithmic cooling
(N$^{{\rm cold}}$).}
\label{15n-relax-table}
\end{table}

\noindent{\bf Magnetization trajectories during compression unitary 
implementation:}\\
The plots of individual spin magnetizations of the ${}^{13}$C-${}^{15}$N-labeled
formamide system at the end of the implementation of each quantum gate in the
quantum circuit of the TSAC algorithm are shown in Fig.~\ref{mag-plot}(b).  The
first three gates comprise the right shift operator, after implementation of
which, the entropy of the ${}^{15}$N spin is decreased, that of the ${}^{13}$C
spin is maximized and that of the ${}^{1}$H spin remains the same. Hence at this
stage, the magnetizations of the ${}^{15}$N, ${}^{13}$C and ${}^{1}$H spins are
halved, zeroed and inverted, respectively.  The sixth gate in the circuit, which
is a Toffoli gate with control on the ${}^{1}$H and ${}^{13}$C spins and target
on the ${}^{15}$N spin has decreased the entropy of the ${}^{15}$N spin and
correspondingly increased its polarization to its final experimental
polarization.  The final CNOT and NOT gates act to decrease the entropy
of the ${}^{13}$C spin and invert the magnetization of the ${}^{1}$H spin.

In summary, the TSAC cooling protocol performs very well in cooling spins with
low initial polarizations such as ${}^{13}$C and ${}^{15}$N spins. The protocol
is robust and experimentally feasible. While maximum possible cooling was not
achieved experimentally for the
${}^{15}$N spin, this could be attributed to several factors including
decoherence, rf inhomogeneities and errors in the calibration of rf pulse
parameters.  The measured T$_1$ relaxation times (in seconds) of all the spins
in the ${}^{13}$C$_2$-labeled glycine system and in the
${}^{13}$C-${}^{15}$N-labeled formamide system are tabulated in
Tables~\ref{13c-relax-table} and \ref{15n-relax-table}, respectively.  The
relaxation dynamics of the algorithmically cooled spins does not change
appreciably after implementation of the TSAC protocol, implying that these
systems retain their good dynamics properties, which are important for quantum
computing.  The gap between the numerically computed upper bounds on the
achievable polarizations for perfect TSAC conditions and the experimentally
achieved polarizations suggests that cooling can be enhanced in systems with
favorable relaxation properties by implementing several cycles of the TSAC
protocol.
\section{Conclusions}
\label{sec4}
A major drawback of all standard HBAC protocols is that the operations for
implementing compression are rather complex, change after every iteration, and
hence require knowledge of the state in each iteration.  A new HBAC protocol was
designed which reaches the cooling limit using a fixed state-independent
operation as the compression step, TSAC method.  We used the TSAC method to
purify selected target qubits in a system containing computational and reset
qubits. We designed an optimal circuit decomposition of the compression unitary
used in the protocol in terms of standard CNOT, Toffoli and NOT gates which are
experimentally viable to implement.  Using an NMR quantum processor, we
experimentally demonstrated the efficacy of the two-sort algorithmic cooling
method in two different systems containing a target ${}^{13}$C and ${}^{15}$N
spin. We obtained large polarization enhancements 
for both spins, which implies an appreciable
decrease in the corresponding spin temperatures.  We achieved significant
cooling of the ${}^{15}$N spin, with a polarization enhancement by a factor of
$\approx 5.95$ which implies cooling down to a 
temperature $\approx 51~K$.  Since
${}^{15}$N has a very low polarization at room temperatures, our work has
important implications for enhancing the signal-to-noise ratios, thereby
reducing experimental acquisition times for ${15}$N-labeled biomolecules such as
peptides and proteins.

Since heat-bath algorithmic cooling methods reduce the spin entropy and achieve
spin cooling (and enhanced spin polarization) without the need for physical
cooling of the system, they are safe and robust and have 
immense applications for
{\em in vivo} NMR spectroscopy of slow metabolic processes.  The two-sort
heat-bath algorithmic cooling method is general and experimentally feasible, and
can be used to cool qubits in other quantum processing devices as well.  Future
work in this direction includes using the two-sort algorithm to enhance cooling
of target qubits in larger-qubit registers and to investigate the performance of
this algorithm and the attainable cooling limit in the presence of realistic
noise.  

\begin{acknowledgments}
All experiments were performed on a Bruker Avance-III 600 MHz NMR spectrometer
at the NMR Research Facility at IISER Mohali.  K.~S. thanks Akshay, Dileep and
Mamta for useful discussions.  K.~S. acknowledges financial support from the
Prime Minister’s Research Fellowship(PMRF) scheme of the Government of India.
Arvind acknowledges funding from the Department of Science and Technology (DST),
India, Grant No:DST/ICPS/QuST/Theme-1/2019/Q-68.  K.D.  acknowledges funding
from the Department of Science and Technology (DST), India, Grant
No:DST/ICPS/QuST/Theme-2/2019/Q-74.
\end{acknowledgments}

\begin{thebibliography}{43}%
\makeatletter
\providecommand \@ifxundefined [1]{%
 \@ifx{#1\undefined}
}%
\providecommand \@ifnum [1]{%
 \ifnum #1\expandafter \@firstoftwo
 \else \expandafter \@secondoftwo
 \fi
}%
\providecommand \@ifx [1]{%
 \ifx #1\expandafter \@firstoftwo
 \else \expandafter \@secondoftwo
 \fi
}%
\providecommand \natexlab [1]{#1}%
\providecommand \enquote  [1]{``#1''}%
\providecommand \bibnamefont  [1]{#1}%
\providecommand \bibfnamefont [1]{#1}%
\providecommand \citenamefont [1]{#1}%
\providecommand \href@noop [0]{\@secondoftwo}%
\providecommand \href [0]{\begingroup \@sanitize@url \@href}%
\providecommand \@href[1]{\@@startlink{#1}\@@href}%
\providecommand \@@href[1]{\endgroup#1\@@endlink}%
\providecommand \@sanitize@url [0]{\catcode `\\12\catcode `\$12\catcode
  `\&12\catcode `\#12\catcode `\^12\catcode `\_12\catcode `\%12\relax}%
\providecommand \@@startlink[1]{}%
\providecommand \@@endlink[0]{}%
\providecommand \url  [0]{\begingroup\@sanitize@url \@url }%
\providecommand \@url [1]{\endgroup\@href {#1}{\urlprefix }}%
\providecommand \urlprefix  [0]{URL }%
\providecommand \Eprint [0]{\href }%
\providecommand \doibase [0]{http://dx.doi.org/}%
\providecommand \selectlanguage [0]{\@gobble}%
\providecommand \bibinfo  [0]{\@secondoftwo}%
\providecommand \bibfield  [0]{\@secondoftwo}%
\providecommand \translation [1]{[#1]}%
\providecommand \BibitemOpen [0]{}%
\providecommand \bibitemStop [0]{}%
\providecommand \bibitemNoStop [0]{.\EOS\space}%
\providecommand \EOS [0]{\spacefactor3000\relax}%
\providecommand \BibitemShut  [1]{\csname bibitem#1\endcsname}%
\let\auto@bib@innerbib\@empty
\bibitem [{\citenamefont {Raeisi}\ \emph {et~al.}(2019)\citenamefont {Raeisi},
  \citenamefont {Kieferov\'a},\ and\ \citenamefont {Mosca}}]{Sadegh}%
  \BibitemOpen
  \bibfield  {author} {\bibinfo {author} {\bibfnamefont {S.}~\bibnamefont
  {Raeisi}}, \bibinfo {author} {\bibfnamefont {M.}~\bibnamefont {Kieferov\'a}},
  \ and\ \bibinfo {author} {\bibfnamefont {M.}~\bibnamefont {Mosca}},\ }\href
  {\doibase 10.1103/PhysRevLett.122.220501} {\bibfield  {journal} {\bibinfo
  {journal} {Phys. Rev. Lett.}\ }\textbf {\bibinfo {volume} {122}},\ \bibinfo
  {pages} {220501} (\bibinfo {year} {2019})}\BibitemShut {NoStop}%
\bibitem [{\citenamefont {Nielsen}\ and\ \citenamefont
  {Chuang}(2010)}]{nielsen}%
  \BibitemOpen
  \bibfield  {author} {\bibinfo {author} {\bibfnamefont {M.~A.}\ \bibnamefont
  {Nielsen}}\ and\ \bibinfo {author} {\bibfnamefont {I.~L.}\ \bibnamefont
  {Chuang}},\ }\href {\doibase 10.1017/CBO9780511976667} {\emph {\bibinfo
  {title} {Quantum Computation and Quantum Information}}}\ (\bibinfo
  {publisher} {Cambridge University Press},\ \bibinfo {year}
  {2010})\BibitemShut {NoStop}%
\bibitem [{\citenamefont {Shor}(1997)}]{shor}%
  \BibitemOpen
  \bibfield  {author} {\bibinfo {author} {\bibfnamefont {P.~W.}\ \bibnamefont
  {Shor}},\ }\href {\doibase 10.1137/S0097539795293172} {\bibfield  {journal}
  {\bibinfo  {journal} {SIAM Journal on Computing}\ }\textbf {\bibinfo {volume}
  {26}},\ \bibinfo {pages} {1484} (\bibinfo {year} {1997})}\BibitemShut
  {NoStop}%
\bibitem [{\citenamefont {Harrow}\ \emph {et~al.}(2009)\citenamefont {Harrow},
  \citenamefont {Hassidim},\ and\ \citenamefont {Lloyd}}]{harrow}%
  \BibitemOpen
  \bibfield  {author} {\bibinfo {author} {\bibfnamefont {A.~W.}\ \bibnamefont
  {Harrow}}, \bibinfo {author} {\bibfnamefont {A.}~\bibnamefont {Hassidim}}, \
  and\ \bibinfo {author} {\bibfnamefont {S.}~\bibnamefont {Lloyd}},\ }\href
  {\doibase 10.1103/PhysRevLett.103.150502} {\bibfield  {journal} {\bibinfo
  {journal} {Phys. Rev. Lett.}\ }\textbf {\bibinfo {volume} {103}},\ \bibinfo
  {pages} {150502} (\bibinfo {year} {2009})}\BibitemShut {NoStop}%
\bibitem [{\citenamefont {Wang}\ \emph {et~al.}(2013)\citenamefont {Wang},
  \citenamefont {Ghobadi}, \citenamefont {Raeisi},\ and\ \citenamefont
  {Simon}}]{quanteff}%
  \BibitemOpen
  \bibfield  {author} {\bibinfo {author} {\bibfnamefont {T.}~\bibnamefont
  {Wang}}, \bibinfo {author} {\bibfnamefont {R.}~\bibnamefont {Ghobadi}},
  \bibinfo {author} {\bibfnamefont {S.}~\bibnamefont {Raeisi}}, \ and\ \bibinfo
  {author} {\bibfnamefont {C.}~\bibnamefont {Simon}},\ }\href {\doibase
  10.1103/PhysRevA.88.062114} {\bibfield  {journal} {\bibinfo  {journal} {Phys.
  Rev. A}\ }\textbf {\bibinfo {volume} {88}},\ \bibinfo {pages} {062114}
  (\bibinfo {year} {2013})}\BibitemShut {NoStop}%
\bibitem [{\citenamefont {Preskill}(1998)}]{qec}%
  \BibitemOpen
  \bibfield  {author} {\bibinfo {author} {\bibfnamefont {J.}~\bibnamefont
  {Preskill}},\ }\href {\doibase 10.1098/rspa.1998.0167} {\bibfield  {journal}
  {\bibinfo  {journal} {Proceedings of the Royal Society of London. Series A:
  Mathematical, Physical and Engineering Sciences}\ }\textbf {\bibinfo {volume}
  {454}},\ \bibinfo {pages} {385} (\bibinfo {year} {1998})}\BibitemShut
  {NoStop}%
\bibitem [{\citenamefont {Cory}\ \emph {et~al.}(2000)\citenamefont {Cory},
  \citenamefont {Laflamme}, \citenamefont {Knill}, \citenamefont {Viola},
  \citenamefont {Havel}, \citenamefont {Boulant}, \citenamefont {Boutis},
  \citenamefont {Fortunato}, \citenamefont {Lloyd}, \citenamefont {Martinez},
  \citenamefont {Negrevergne}, \citenamefont {Pravia}, \citenamefont {Sharf},
  \citenamefont {Teklemariam}, \citenamefont {Weinstein},\ and\ \citenamefont
  {Zurek}}]{Cory2000}%
  \BibitemOpen
  \bibfield  {author} {\bibinfo {author} {\bibfnamefont {D.~G.}\ \bibnamefont
  {Cory}}, \bibinfo {author} {\bibfnamefont {R.}~\bibnamefont {Laflamme}},
  \bibinfo {author} {\bibfnamefont {E.}~\bibnamefont {Knill}}, \bibinfo
  {author} {\bibfnamefont {L.}~\bibnamefont {Viola}}, \bibinfo {author}
  {\bibfnamefont {T.~F.}\ \bibnamefont {Havel}}, \bibinfo {author}
  {\bibfnamefont {N.}~\bibnamefont {Boulant}}, \bibinfo {author} {\bibfnamefont
  {G.}~\bibnamefont {Boutis}}, \bibinfo {author} {\bibfnamefont
  {E.}~\bibnamefont {Fortunato}}, \bibinfo {author} {\bibfnamefont
  {S.}~\bibnamefont {Lloyd}}, \bibinfo {author} {\bibfnamefont
  {R.}~\bibnamefont {Martinez}}, \bibinfo {author} {\bibfnamefont
  {C.}~\bibnamefont {Negrevergne}}, \bibinfo {author} {\bibfnamefont
  {M.}~\bibnamefont {Pravia}}, \bibinfo {author} {\bibfnamefont
  {Y.}~\bibnamefont {Sharf}}, \bibinfo {author} {\bibfnamefont
  {G.}~\bibnamefont {Teklemariam}}, \bibinfo {author} {\bibfnamefont {Y.~S.}\
  \bibnamefont {Weinstein}}, \ and\ \bibinfo {author} {\bibfnamefont {W.~H.}\
  \bibnamefont {Zurek}},\ }\href {\doibase
  10.1002/1521-3978(200009)48:9/11<875::AID-PROP875>3.0.CO;2-V} {\bibfield
  {journal} {\bibinfo  {journal} {Fortschritte der Physik}\ }\textbf {\bibinfo
  {volume} {48}},\ \bibinfo {pages} {875} (\bibinfo {year} {2000})}\BibitemShut
  {NoStop}%
\bibitem [{\citenamefont {Vandersypen}\ and\ \citenamefont
  {Chuang}(2005)}]{vandersypen}%
  \BibitemOpen
  \bibfield  {author} {\bibinfo {author} {\bibfnamefont {L.~M.~K.}\
  \bibnamefont {Vandersypen}}\ and\ \bibinfo {author} {\bibfnamefont {I.~L.}\
  \bibnamefont {Chuang}},\ }\href {\doibase 10.1103/RevModPhys.76.1037}
  {\bibfield  {journal} {\bibinfo  {journal} {Rev. Mod. Phys.}\ }\textbf
  {\bibinfo {volume} {76}},\ \bibinfo {pages} {1037} (\bibinfo {year}
  {2005})}\BibitemShut {NoStop}%
\bibitem [{\citenamefont {Dorai}\ \emph {et~al.}(2000)\citenamefont {Dorai},
  \citenamefont {Mahesh}, \citenamefont {Arvind},\ and\ \citenamefont
  {Kumar}}]{KD-current-science}%
  \BibitemOpen
  \bibfield  {author} {\bibinfo {author} {\bibfnamefont {K.}~\bibnamefont
  {Dorai}}, \bibinfo {author} {\bibfnamefont {T.~S.}\ \bibnamefont {Mahesh}},
  \bibinfo {author} {\bibnamefont {Arvind}}, \ and\ \bibinfo {author}
  {\bibfnamefont {A.}~\bibnamefont {Kumar}},\ }\href@noop {} {\bibfield
  {journal} {\bibinfo  {journal} {Current Science}\ }\textbf {\bibinfo {volume}
  {79}},\ \bibinfo {pages} {1447} (\bibinfo {year} {2000})}\BibitemShut
  {NoStop}%
\bibitem [{\citenamefont {{Divincenzo}}(2000)}]{divincenzo}%
  \BibitemOpen
  \bibfield  {author} {\bibinfo {author} {\bibfnamefont {D.~P.}\ \bibnamefont
  {{Divincenzo}}},\ }\href {\doibase
  10.1002/1521-3978(200009)48:9/11<771::AID-PROP771>3.0.CO;2-E} {\bibfield
  {journal} {\bibinfo  {journal} {Fortschritte der Physik}\ }\textbf {\bibinfo
  {volume} {48}},\ \bibinfo {pages} {771} (\bibinfo {year} {2000})}\BibitemShut
  {NoStop}%
\bibitem [{\citenamefont {Linden}\ and\ \citenamefont
  {Popescu}(2001)}]{Linden}%
  \BibitemOpen
  \bibfield  {author} {\bibinfo {author} {\bibfnamefont {N.}~\bibnamefont
  {Linden}}\ and\ \bibinfo {author} {\bibfnamefont {S.}~\bibnamefont
  {Popescu}},\ }\href {\doibase 10.1103/PhysRevLett.87.047901} {\bibfield
  {journal} {\bibinfo  {journal} {Phys. Rev. Lett.}\ }\textbf {\bibinfo
  {volume} {87}},\ \bibinfo {pages} {047901} (\bibinfo {year}
  {2001})}\BibitemShut {NoStop}%
\bibitem [{\citenamefont {Schulman}\ and\ \citenamefont
  {Vazirani}(1999)}]{Schulman}%
  \BibitemOpen
  \bibfield  {author} {\bibinfo {author} {\bibfnamefont {L.~J.}\ \bibnamefont
  {Schulman}}\ and\ \bibinfo {author} {\bibfnamefont {U.~V.}\ \bibnamefont
  {Vazirani}},\ }in\ \href@noop {} {\emph {\bibinfo {booktitle} {IN 31ST
  STOC}}}\ (\bibinfo {year} {1999})\ pp.\ \bibinfo {pages}
  {322--329}\BibitemShut {NoStop}%
\bibitem [{\citenamefont {Ardenkjaer-Larsen}\ \emph {et~al.}(2003)\citenamefont
  {Ardenkjaer-Larsen}, \citenamefont {Fridlund}, \citenamefont {Gram},
  \citenamefont {Hansson}, \citenamefont {Hansson}, \citenamefont {Lerche},
  \citenamefont {Servin}, \citenamefont {Thaning},\ and\ \citenamefont
  {Golman}}]{dnp}%
  \BibitemOpen
  \bibfield  {author} {\bibinfo {author} {\bibfnamefont {J.~H.}\ \bibnamefont
  {Ardenkjaer-Larsen}}, \bibinfo {author} {\bibfnamefont {B.}~\bibnamefont
  {Fridlund}}, \bibinfo {author} {\bibfnamefont {A.}~\bibnamefont {Gram}},
  \bibinfo {author} {\bibfnamefont {G.}~\bibnamefont {Hansson}}, \bibinfo
  {author} {\bibfnamefont {L.}~\bibnamefont {Hansson}}, \bibinfo {author}
  {\bibfnamefont {M.~H.}\ \bibnamefont {Lerche}}, \bibinfo {author}
  {\bibfnamefont {R.}~\bibnamefont {Servin}}, \bibinfo {author} {\bibfnamefont
  {M.}~\bibnamefont {Thaning}}, \ and\ \bibinfo {author} {\bibfnamefont
  {K.}~\bibnamefont {Golman}},\ }\href {\doibase 10.1073/pnas.1733835100}
  {\bibfield  {journal} {\bibinfo  {journal} {Proceedings of the National
  Academy of Sciences}\ }\textbf {\bibinfo {volume} {100}},\ \bibinfo {pages}
  {10158} (\bibinfo {year} {2003})}\BibitemShut {NoStop}%
\bibitem [{\citenamefont {Bhattacharya}\ \emph {et~al.}(2005)\citenamefont
  {Bhattacharya}, \citenamefont {Harris}, \citenamefont {Lin}, \citenamefont
  {Mansson}, \citenamefont {Norton}, \citenamefont {Perman}, \citenamefont
  {Weitekamp},\ and\ \citenamefont {Ross}}]{phip}%
  \BibitemOpen
  \bibfield  {author} {\bibinfo {author} {\bibfnamefont {P.}~\bibnamefont
  {Bhattacharya}}, \bibinfo {author} {\bibfnamefont {K.}~\bibnamefont
  {Harris}}, \bibinfo {author} {\bibfnamefont {A.~P.}\ \bibnamefont {Lin}},
  \bibinfo {author} {\bibfnamefont {M.}~\bibnamefont {Mansson}}, \bibinfo
  {author} {\bibfnamefont {V.~A.}\ \bibnamefont {Norton}}, \bibinfo {author}
  {\bibfnamefont {W.~H.}\ \bibnamefont {Perman}}, \bibinfo {author}
  {\bibfnamefont {D.~P.}\ \bibnamefont {Weitekamp}}, \ and\ \bibinfo {author}
  {\bibfnamefont {B.~D.}\ \bibnamefont {Ross}},\ }\href@noop {} {\bibfield
  {journal} {\bibinfo  {journal} {MAGMA}\ }\textbf {\bibinfo {volume} {18}},\
  \bibinfo {pages} {245} (\bibinfo {year} {2005})}\BibitemShut {NoStop}%
\bibitem [{\citenamefont {Oros}\ and\ \citenamefont
  {Shah}(2004)}]{opticalpump}%
  \BibitemOpen
  \bibfield  {author} {\bibinfo {author} {\bibfnamefont {A.-M.}\ \bibnamefont
  {Oros}}\ and\ \bibinfo {author} {\bibfnamefont {N.~J.}\ \bibnamefont
  {Shah}},\ }\href {\doibase 10.1088/0031-9155/49/20/R01} {\bibfield  {journal}
  {\bibinfo  {journal} {Physics in Medicine and Biology}\ }\textbf {\bibinfo
  {volume} {49}},\ \bibinfo {pages} {R105} (\bibinfo {year}
  {2004})}\BibitemShut {NoStop}%
\bibitem [{\citenamefont {Cover}\ and\ \citenamefont
  {Thomas}(2006)}]{Cover2006}%
  \BibitemOpen
  \bibfield  {author} {\bibinfo {author} {\bibfnamefont {T.~M.}\ \bibnamefont
  {Cover}}\ and\ \bibinfo {author} {\bibfnamefont {J.~A.}\ \bibnamefont
  {Thomas}},\ }\href {\doibase 10.1002/047174882X} {\emph {\bibinfo {title}
  {Elements of Information Theory}}}\ (\bibinfo  {publisher}
  {Wiley-Interscience},\ \bibinfo {year} {2006})\BibitemShut {NoStop}%
\bibitem [{\citenamefont {Boykin}\ \emph {et~al.}(2002)\citenamefont {Boykin},
  \citenamefont {Mor}, \citenamefont {Roychowdhury}, \citenamefont {Vatan},\
  and\ \citenamefont {Vrijen}}]{Boykin}%
  \BibitemOpen
  \bibfield  {author} {\bibinfo {author} {\bibfnamefont {P.~O.}\ \bibnamefont
  {Boykin}}, \bibinfo {author} {\bibfnamefont {T.}~\bibnamefont {Mor}},
  \bibinfo {author} {\bibfnamefont {V.}~\bibnamefont {Roychowdhury}}, \bibinfo
  {author} {\bibfnamefont {F.}~\bibnamefont {Vatan}}, \ and\ \bibinfo {author}
  {\bibfnamefont {R.}~\bibnamefont {Vrijen}},\ }\href {\doibase
  10.1073/pnas.241641898} {\bibfield  {journal} {\bibinfo  {journal}
  {Proceedings of the National Academy of Sciences}\ }\textbf {\bibinfo
  {volume} {99}},\ \bibinfo {pages} {3388} (\bibinfo {year}
  {2002})}\BibitemShut {NoStop}%
\bibitem [{\citenamefont {Fernandez}\ \emph {et~al.}(2004)\citenamefont
  {Fernandez}, \citenamefont {Lloyd}, \citenamefont {Mor},\ and\ \citenamefont
  {Roychowdhury}}]{jose}%
  \BibitemOpen
  \bibfield  {author} {\bibinfo {author} {\bibfnamefont {J.~M.}\ \bibnamefont
  {Fernandez}}, \bibinfo {author} {\bibfnamefont {S.}~\bibnamefont {Lloyd}},
  \bibinfo {author} {\bibfnamefont {T.}~\bibnamefont {Mor}}, \ and\ \bibinfo
  {author} {\bibfnamefont {V.}~\bibnamefont {Roychowdhury}},\ }\href {\doibase
  10.1142/S0219749904000419} {\bibfield  {journal} {\bibinfo  {journal}
  {International Journal of Quantum Information}\ }\textbf {\bibinfo {volume}
  {02}},\ \bibinfo {pages} {461} (\bibinfo {year} {2004})}\BibitemShut
  {NoStop}%
\bibitem [{\citenamefont {Schulman}\ \emph {et~al.}(2007)\citenamefont
  {Schulman}, \citenamefont {Mor},\ and\ \citenamefont
  {Weinstein}}]{schulman3}%
  \BibitemOpen
  \bibfield  {author} {\bibinfo {author} {\bibfnamefont {L.~J.}\ \bibnamefont
  {Schulman}}, \bibinfo {author} {\bibfnamefont {T.}~\bibnamefont {Mor}}, \
  and\ \bibinfo {author} {\bibfnamefont {Y.}~\bibnamefont {Weinstein}},\ }\href
  {\doibase 10.1137/050666023} {\bibfield  {journal} {\bibinfo  {journal} {SIAM
  Journal on Computing}\ }\textbf {\bibinfo {volume} {36}},\ \bibinfo {pages}
  {1729} (\bibinfo {year} {2007})}\BibitemShut {NoStop}%
\bibitem [{\citenamefont {Schulman}\ \emph {et~al.}(2005)\citenamefont
  {Schulman}, \citenamefont {Mor},\ and\ \citenamefont
  {Weinstein}}]{Schulman2}%
  \BibitemOpen
  \bibfield  {author} {\bibinfo {author} {\bibfnamefont {L.~J.}\ \bibnamefont
  {Schulman}}, \bibinfo {author} {\bibfnamefont {T.}~\bibnamefont {Mor}}, \
  and\ \bibinfo {author} {\bibfnamefont {Y.}~\bibnamefont {Weinstein}},\ }\href
  {\doibase 10.1103/PhysRevLett.94.120501} {\bibfield  {journal} {\bibinfo
  {journal} {Phys. Rev. Lett.}\ }\textbf {\bibinfo {volume} {94}},\ \bibinfo
  {pages} {120501} (\bibinfo {year} {2005})}\BibitemShut {NoStop}%
\bibitem [{\citenamefont {Elias}\ \emph
  {et~al.}(2011{\natexlab{a}})\citenamefont {Elias}, \citenamefont {Mor},\ and\
  \citenamefont {Weinstein}}]{elias}%
  \BibitemOpen
  \bibfield  {author} {\bibinfo {author} {\bibfnamefont {Y.}~\bibnamefont
  {Elias}}, \bibinfo {author} {\bibfnamefont {T.}~\bibnamefont {Mor}}, \ and\
  \bibinfo {author} {\bibfnamefont {Y.}~\bibnamefont {Weinstein}},\ }\href
  {\doibase 10.1103/PhysRevA.83.042340} {\bibfield  {journal} {\bibinfo
  {journal} {Phys. Rev. A}\ }\textbf {\bibinfo {volume} {83}},\ \bibinfo
  {pages} {042340} (\bibinfo {year} {2011}{\natexlab{a}})}\BibitemShut
  {NoStop}%
\bibitem [{\citenamefont {Raeisi}\ and\ \citenamefont {Mosca}(2015)}]{asym}%
  \BibitemOpen
  \bibfield  {author} {\bibinfo {author} {\bibfnamefont {S.}~\bibnamefont
  {Raeisi}}\ and\ \bibinfo {author} {\bibfnamefont {M.}~\bibnamefont {Mosca}},\
  }\href {\doibase 10.1103/PhysRevLett.114.100404} {\bibfield  {journal}
  {\bibinfo  {journal} {Phys. Rev. Lett.}\ }\textbf {\bibinfo {volume} {114}},\
  \bibinfo {pages} {100404} (\bibinfo {year} {2015})}\BibitemShut {NoStop}%
\bibitem [{\citenamefont {Rodr\'{\i}guez-Briones}\ and\ \citenamefont
  {Laflamme}(2016)}]{pollimit}%
  \BibitemOpen
  \bibfield  {author} {\bibinfo {author} {\bibfnamefont {N.~A.}\ \bibnamefont
  {Rodr\'{\i}guez-Briones}}\ and\ \bibinfo {author} {\bibfnamefont
  {R.}~\bibnamefont {Laflamme}},\ }\href {\doibase
  10.1103/PhysRevLett.116.170501} {\bibfield  {journal} {\bibinfo  {journal}
  {Phys. Rev. Lett.}\ }\textbf {\bibinfo {volume} {116}},\ \bibinfo {pages}
  {170501} (\bibinfo {year} {2016})}\BibitemShut {NoStop}%
\bibitem [{\citenamefont {Alhambra}\ \emph {et~al.}(2019)\citenamefont
  {Alhambra}, \citenamefont {Lostaglio},\ and\ \citenamefont
  {Perry}}]{AlhambraHBAC}%
  \BibitemOpen
  \bibfield  {author} {\bibinfo {author} {\bibfnamefont {{\'{A}}.~M.}\
  \bibnamefont {Alhambra}}, \bibinfo {author} {\bibfnamefont {M.}~\bibnamefont
  {Lostaglio}}, \ and\ \bibinfo {author} {\bibfnamefont {C.}~\bibnamefont
  {Perry}},\ }\href {\doibase 10.22331/q-2019-09-23-188} {\bibfield  {journal}
  {\bibinfo  {journal} {{Quantum}}\ }\textbf {\bibinfo {volume} {3}},\ \bibinfo
  {pages} {188} (\bibinfo {year} {2019})}\BibitemShut {NoStop}%
\bibitem [{\citenamefont {Raeisi}(2021)}]{Raeisi_No-go}%
  \BibitemOpen
  \bibfield  {author} {\bibinfo {author} {\bibfnamefont {S.}~\bibnamefont
  {Raeisi}},\ }\href {\doibase 10.1103/PhysRevA.103.062424} {\bibfield
  {journal} {\bibinfo  {journal} {Phys. Rev. A}\ }\textbf {\bibinfo {volume}
  {103}},\ \bibinfo {pages} {062424} (\bibinfo {year} {2021})}\BibitemShut
  {NoStop}%
\bibitem [{\citenamefont {Farahmand}\ \emph {et~al.}(2022)\citenamefont
  {Farahmand}, \citenamefont {Aghaei~Saem},\ and\ \citenamefont
  {Raeisi}}]{Farahmand}%
  \BibitemOpen
  \bibfield  {author} {\bibinfo {author} {\bibfnamefont {Z.}~\bibnamefont
  {Farahmand}}, \bibinfo {author} {\bibfnamefont {R.}~\bibnamefont
  {Aghaei~Saem}}, \ and\ \bibinfo {author} {\bibfnamefont {S.}~\bibnamefont
  {Raeisi}},\ }\href {\doibase 10.1103/PhysRevA.105.022418} {\bibfield
  {journal} {\bibinfo  {journal} {Phys. Rev. A}\ }\textbf {\bibinfo {volume}
  {105}},\ \bibinfo {pages} {022418} (\bibinfo {year} {2022})}\BibitemShut
  {NoStop}%
\bibitem [{\citenamefont {Baugh}\ \emph {et~al.}(2005)\citenamefont {Baugh},
  \citenamefont {Moussa}, \citenamefont {Ryan}, \citenamefont {Nayak},\ and\
  \citenamefont {Laflamme}}]{Baugh}%
  \BibitemOpen
  \bibfield  {author} {\bibinfo {author} {\bibfnamefont {J.}~\bibnamefont
  {Baugh}}, \bibinfo {author} {\bibfnamefont {O.}~\bibnamefont {Moussa}},
  \bibinfo {author} {\bibfnamefont {C.~A.}\ \bibnamefont {Ryan}}, \bibinfo
  {author} {\bibfnamefont {A.}~\bibnamefont {Nayak}}, \ and\ \bibinfo {author}
  {\bibfnamefont {R.}~\bibnamefont {Laflamme}},\ }\href {\doibase
  10.1038/nature04272} {\bibfield  {journal} {\bibinfo  {journal} {Nature}\
  }\textbf {\bibinfo {volume} {438}},\ \bibinfo {pages} {470} (\bibinfo {year}
  {2005})}\BibitemShut {NoStop}%
\bibitem [{\citenamefont {Fernandez}\ \emph {et~al.}(2005)\citenamefont
  {Fernandez}, \citenamefont {Mor},\ and\ \citenamefont
  {Weinstein}}]{paramagnet}%
  \BibitemOpen
  \bibfield  {author} {\bibinfo {author} {\bibfnamefont {J.~M.}\ \bibnamefont
  {Fernandez}}, \bibinfo {author} {\bibfnamefont {T.}~\bibnamefont {Mor}}, \
  and\ \bibinfo {author} {\bibfnamefont {Y.}~\bibnamefont {Weinstein}},\ }\href
  {\doibase 10.1142/S0219749905000888} {\bibfield  {journal} {\bibinfo
  {journal} {International Journal of Quantum Information}\ }\textbf {\bibinfo
  {volume} {03}},\ \bibinfo {pages} {281} (\bibinfo {year} {2005})}\BibitemShut
  {NoStop}%
\bibitem [{\citenamefont {Elias}\ \emph
  {et~al.}(2011{\natexlab{b}})\citenamefont {Elias}, \citenamefont {Gilboa},
  \citenamefont {Mor},\ and\ \citenamefont {Weinstein}}]{amino}%
  \BibitemOpen
  \bibfield  {author} {\bibinfo {author} {\bibfnamefont {Y.}~\bibnamefont
  {Elias}}, \bibinfo {author} {\bibfnamefont {H.}~\bibnamefont {Gilboa}},
  \bibinfo {author} {\bibfnamefont {T.}~\bibnamefont {Mor}}, \ and\ \bibinfo
  {author} {\bibfnamefont {Y.}~\bibnamefont {Weinstein}},\ }\href {\doibase
  https://doi.org/10.1016/j.cplett.2011.10.039} {\bibfield  {journal} {\bibinfo
   {journal} {Chemical Physics Letters}\ }\textbf {\bibinfo {volume} {517}},\
  \bibinfo {pages} {126} (\bibinfo {year} {2011}{\natexlab{b}})}\BibitemShut
  {NoStop}%
\bibitem [{\citenamefont {Brassard}\ \emph {et~al.}(2014)\citenamefont
  {Brassard}, \citenamefont {Elias}, \citenamefont {Fernandez}, \citenamefont
  {Gilboa}, \citenamefont {Jones}, \citenamefont {Mor}, \citenamefont
  {Weinstein},\ and\ \citenamefont {Xiao}}]{Brassard}%
  \BibitemOpen
  \bibfield  {author} {\bibinfo {author} {\bibfnamefont {G.}~\bibnamefont
  {Brassard}}, \bibinfo {author} {\bibfnamefont {Y.}~\bibnamefont {Elias}},
  \bibinfo {author} {\bibfnamefont {J.~M.}\ \bibnamefont {Fernandez}}, \bibinfo
  {author} {\bibfnamefont {H.}~\bibnamefont {Gilboa}}, \bibinfo {author}
  {\bibfnamefont {J.~A.}\ \bibnamefont {Jones}}, \bibinfo {author}
  {\bibfnamefont {T.}~\bibnamefont {Mor}}, \bibinfo {author} {\bibfnamefont
  {Y.}~\bibnamefont {Weinstein}}, \ and\ \bibinfo {author} {\bibfnamefont
  {L.}~\bibnamefont {Xiao}},\ }\href {\doibase 10.1140/epjp/i2014-14266-0}
  {\bibfield  {journal} {\bibinfo  {journal} {The European Physical Journal
  Plus}\ }\textbf {\bibinfo {volume} {129}},\ \bibinfo {pages} {266} (\bibinfo
  {year} {2014})}\BibitemShut {NoStop}%
\bibitem [{\citenamefont {Atia}\ \emph {et~al.}(2016)\citenamefont {Atia},
  \citenamefont {Elias}, \citenamefont {Mor},\ and\ \citenamefont
  {Weinstein}}]{yosi}%
  \BibitemOpen
  \bibfield  {author} {\bibinfo {author} {\bibfnamefont {Y.}~\bibnamefont
  {Atia}}, \bibinfo {author} {\bibfnamefont {Y.}~\bibnamefont {Elias}},
  \bibinfo {author} {\bibfnamefont {T.}~\bibnamefont {Mor}}, \ and\ \bibinfo
  {author} {\bibfnamefont {Y.}~\bibnamefont {Weinstein}},\ }\href {\doibase
  10.1103/PhysRevA.93.012325} {\bibfield  {journal} {\bibinfo  {journal} {Phys.
  Rev. A}\ }\textbf {\bibinfo {volume} {93}},\ \bibinfo {pages} {012325}
  (\bibinfo {year} {2016})}\BibitemShut {NoStop}%
\bibitem [{\citenamefont {Pande}\ \emph {et~al.}(2017)\citenamefont {Pande},
  \citenamefont {Bhole}, \citenamefont {Khurana},\ and\ \citenamefont
  {Mahesh}}]{tsm}%
  \BibitemOpen
  \bibfield  {author} {\bibinfo {author} {\bibfnamefont {V.~R.}\ \bibnamefont
  {Pande}}, \bibinfo {author} {\bibfnamefont {G.}~\bibnamefont {Bhole}},
  \bibinfo {author} {\bibfnamefont {D.}~\bibnamefont {Khurana}}, \ and\
  \bibinfo {author} {\bibfnamefont {T.~S.}\ \bibnamefont {Mahesh}},\ }\href
  {\doibase 10.1103/PhysRevA.96.012330} {\bibfield  {journal} {\bibinfo
  {journal} {Phys. Rev. A}\ }\textbf {\bibinfo {volume} {96}},\ \bibinfo
  {pages} {012330} (\bibinfo {year} {2017})}\BibitemShut {NoStop}%
\bibitem [{\citenamefont {Ryan}\ \emph {et~al.}(2008)\citenamefont {Ryan},
  \citenamefont {Moussa}, \citenamefont {Baugh},\ and\ \citenamefont
  {Laflamme}}]{laflamme_multiple}%
  \BibitemOpen
  \bibfield  {author} {\bibinfo {author} {\bibfnamefont {C.~A.}\ \bibnamefont
  {Ryan}}, \bibinfo {author} {\bibfnamefont {O.}~\bibnamefont {Moussa}},
  \bibinfo {author} {\bibfnamefont {J.}~\bibnamefont {Baugh}}, \ and\ \bibinfo
  {author} {\bibfnamefont {R.}~\bibnamefont {Laflamme}},\ }\href {\doibase
  10.1103/PhysRevLett.100.140501} {\bibfield  {journal} {\bibinfo  {journal}
  {Phys. Rev. Lett.}\ }\textbf {\bibinfo {volume} {100}},\ \bibinfo {pages}
  {140501} (\bibinfo {year} {2008})}\BibitemShut {NoStop}%
\bibitem [{\citenamefont {K\"ose}\ \emph {et~al.}(2019)\citenamefont {K\"ose},
  \citenamefont {\ifmmode~\mbox{\c{C}}\else \c{C}\fi{}akmak}, \citenamefont
  {Gen\ifmmode~\mbox{\c{c}}\else \c{c}\fi{}ten}, \citenamefont {Kominis},\ and\
  \citenamefont {M\"ustecapl\ifmmode \imath \else \i
  \fi{}o\ifmmode~\breve{g}\else \u{g}\fi{}lu}}]{algoQHE}%
  \BibitemOpen
  \bibfield  {author} {\bibinfo {author} {\bibfnamefont {E.}~\bibnamefont
  {K\"ose}}, \bibinfo {author} {\bibfnamefont {S.~m.~c.}\ \bibnamefont
  {\ifmmode~\mbox{\c{C}}\else \c{C}\fi{}akmak}}, \bibinfo {author}
  {\bibfnamefont {A.}~\bibnamefont {Gen\ifmmode~\mbox{\c{c}}\else
  \c{c}\fi{}ten}}, \bibinfo {author} {\bibfnamefont {I.~K.}\ \bibnamefont
  {Kominis}}, \ and\ \bibinfo {author} {\bibfnamefont {O.~E.}\ \bibnamefont
  {M\"ustecapl\ifmmode \imath \else \i \fi{}o\ifmmode~\breve{g}\else
  \u{g}\fi{}lu}},\ }\href {\doibase 10.1103/PhysRevE.100.012109} {\bibfield
  {journal} {\bibinfo  {journal} {Phys. Rev. E}\ }\textbf {\bibinfo {volume}
  {100}},\ \bibinfo {pages} {012109} (\bibinfo {year} {2019})}\BibitemShut
  {NoStop}%
\bibitem [{\citenamefont {Rodríguez-Briones}\ \emph
  {et~al.}(2017)\citenamefont {Rodríguez-Briones}, \citenamefont {Li},
  \citenamefont {Peng}, \citenamefont {Mor}, \citenamefont {Weinstein},\ and\
  \citenamefont {Laflamme}}]{laflamme-njp}%
  \BibitemOpen
  \bibfield  {author} {\bibinfo {author} {\bibfnamefont {N.~A.}\ \bibnamefont
  {Rodríguez-Briones}}, \bibinfo {author} {\bibfnamefont {J.}~\bibnamefont
  {Li}}, \bibinfo {author} {\bibfnamefont {X.}~\bibnamefont {Peng}}, \bibinfo
  {author} {\bibfnamefont {T.}~\bibnamefont {Mor}}, \bibinfo {author}
  {\bibfnamefont {Y.}~\bibnamefont {Weinstein}}, \ and\ \bibinfo {author}
  {\bibfnamefont {R.}~\bibnamefont {Laflamme}},\ }\href {\doibase
  10.1088/1367-2630/aa8fe0} {\bibfield  {journal} {\bibinfo  {journal} {New
  Journal of Physics}\ }\textbf {\bibinfo {volume} {19}},\ \bibinfo {pages}
  {113047} (\bibinfo {year} {2017})}\BibitemShut {NoStop}%
\bibitem [{\citenamefont {Zaiser}\ \emph {et~al.}(2021)\citenamefont {Zaiser},
  \citenamefont {Cheung}, \citenamefont {Yang}, \citenamefont {Dasari},
  \citenamefont {Raeisi},\ and\ \citenamefont {Wrachtrup}}]{Zaiser}%
  \BibitemOpen
  \bibfield  {author} {\bibinfo {author} {\bibfnamefont {S.}~\bibnamefont
  {Zaiser}}, \bibinfo {author} {\bibfnamefont {C.~T.}\ \bibnamefont {Cheung}},
  \bibinfo {author} {\bibfnamefont {S.}~\bibnamefont {Yang}}, \bibinfo {author}
  {\bibfnamefont {D.~B.~R.}\ \bibnamefont {Dasari}}, \bibinfo {author}
  {\bibfnamefont {S.}~\bibnamefont {Raeisi}}, \ and\ \bibinfo {author}
  {\bibfnamefont {J.}~\bibnamefont {Wrachtrup}},\ }\href {\doibase
  10.1038/s41534-021-00408-z} {\bibfield  {journal} {\bibinfo  {journal} {npj
  Quantum Information}\ }\textbf {\bibinfo {volume} {7}},\ \bibinfo {pages}
  {92} (\bibinfo {year} {2021})}\BibitemShut {NoStop}%
\bibitem [{\citenamefont {Dorai}\ and\ \citenamefont
  {Suter}(2005)}]{kd-ijqi-2005}%
  \BibitemOpen
  \bibfield  {author} {\bibinfo {author} {\bibfnamefont {K.}~\bibnamefont
  {Dorai}}\ and\ \bibinfo {author} {\bibfnamefont {D.}~\bibnamefont {Suter}},\
  }\href {\doibase 10.1142/S0219749905000967} {\bibfield  {journal} {\bibinfo
  {journal} {International Journal of Quantum Information}\ }\textbf {\bibinfo
  {volume} {03}},\ \bibinfo {pages} {413} (\bibinfo {year} {2005})}\BibitemShut
  {NoStop}%
\bibitem [{\citenamefont {Sørensen}(1989)}]{SORENSEN}%
  \BibitemOpen
  \bibfield  {author} {\bibinfo {author} {\bibfnamefont {O.~W.}\ \bibnamefont
  {Sørensen}},\ }\href {\doibase https://doi.org/10.1016/0079-6565(89)80006-8}
  {\bibfield  {journal} {\bibinfo  {journal} {Progress in Nuclear Magnetic
  Resonance Spectroscopy}\ }\textbf {\bibinfo {volume} {21}},\ \bibinfo {pages}
  {503} (\bibinfo {year} {1989})}\BibitemShut {NoStop}%
\bibitem [{\citenamefont {Barenco}\ \emph {et~al.}(1995)\citenamefont
  {Barenco}, \citenamefont {Bennett}, \citenamefont {Cleve}, \citenamefont
  {DiVincenzo}, \citenamefont {Margolus}, \citenamefont {Shor}, \citenamefont
  {Sleator}, \citenamefont {Smolin},\ and\ \citenamefont {Weinfurter}}]{gate}%
  \BibitemOpen
  \bibfield  {author} {\bibinfo {author} {\bibfnamefont {A.}~\bibnamefont
  {Barenco}}, \bibinfo {author} {\bibfnamefont {C.~H.}\ \bibnamefont
  {Bennett}}, \bibinfo {author} {\bibfnamefont {R.}~\bibnamefont {Cleve}},
  \bibinfo {author} {\bibfnamefont {D.~P.}\ \bibnamefont {DiVincenzo}},
  \bibinfo {author} {\bibfnamefont {N.}~\bibnamefont {Margolus}}, \bibinfo
  {author} {\bibfnamefont {P.}~\bibnamefont {Shor}}, \bibinfo {author}
  {\bibfnamefont {T.}~\bibnamefont {Sleator}}, \bibinfo {author} {\bibfnamefont
  {J.~A.}\ \bibnamefont {Smolin}}, \ and\ \bibinfo {author} {\bibfnamefont
  {H.}~\bibnamefont {Weinfurter}},\ }\href {\doibase 10.1103/PhysRevA.52.3457}
  {\bibfield  {journal} {\bibinfo  {journal} {Phys. Rev. A}\ }\textbf {\bibinfo
  {volume} {52}},\ \bibinfo {pages} {3457} (\bibinfo {year}
  {1995})}\BibitemShut {NoStop}%
\bibitem [{\citenamefont {Oliveira}\ \emph {et~al.}(2007)\citenamefont
  {Oliveira}, \citenamefont {Bonagamba}, \citenamefont {Sarthour},
  \citenamefont {Freitas},\ and\ \citenamefont {deAzevedo}}]{OLIVEIRA}%
  \BibitemOpen
  \bibfield  {author} {\bibinfo {author} {\bibfnamefont {I.~S.}\ \bibnamefont
  {Oliveira}}, \bibinfo {author} {\bibfnamefont {T.~J.}\ \bibnamefont
  {Bonagamba}}, \bibinfo {author} {\bibfnamefont {R.~S.}\ \bibnamefont
  {Sarthour}}, \bibinfo {author} {\bibfnamefont {J.~C.~C.}\ \bibnamefont
  {Freitas}}, \ and\ \bibinfo {author} {\bibfnamefont {E.~R.}\ \bibnamefont
  {deAzevedo}},\ }\href {\doibase
  https://doi.org/10.1016/B978-044452782-0/50006-3} {\emph {\bibinfo {title}
  {NMR quantum information processing}}}\ (\bibinfo  {publisher} {Elsevier
  Science B.V.},\ \bibinfo {address} {Amsterdam},\ \bibinfo {year}
  {2007})\BibitemShut {NoStop}%
\bibitem [{\citenamefont {Das}\ \emph {et~al.}(2015)\citenamefont {Das},
  \citenamefont {Dogra}, \citenamefont {Dorai},\ and\ \citenamefont
  {Arvind}}]{kd-pra-2015-2}%
  \BibitemOpen
  \bibfield  {author} {\bibinfo {author} {\bibfnamefont {D.}~\bibnamefont
  {Das}}, \bibinfo {author} {\bibfnamefont {S.}~\bibnamefont {Dogra}}, \bibinfo
  {author} {\bibfnamefont {K.}~\bibnamefont {Dorai}}, \ and\ \bibinfo {author}
  {\bibnamefont {Arvind}},\ }\href {\doibase 10.1103/PhysRevA.92.022307}
  {\bibfield  {journal} {\bibinfo  {journal} {Phys. Rev. A}\ }\textbf {\bibinfo
  {volume} {92}},\ \bibinfo {pages} {022307} (\bibinfo {year}
  {2015})}\BibitemShut {NoStop}%
\bibitem [{\citenamefont {Singh}\ \emph
  {et~al.}(2018{\natexlab{a}})\citenamefont {Singh}, \citenamefont {Arvind},\
  and\ \citenamefont {Dorai}}]{kd-pra-2018-1}%
  \BibitemOpen
  \bibfield  {author} {\bibinfo {author} {\bibfnamefont {H.}~\bibnamefont
  {Singh}}, \bibinfo {author} {\bibnamefont {Arvind}}, \ and\ \bibinfo {author}
  {\bibfnamefont {K.}~\bibnamefont {Dorai}},\ }\href {\doibase
  10.1103/PhysRevA.97.022302} {\bibfield  {journal} {\bibinfo  {journal} {Phys.
  Rev. A}\ }\textbf {\bibinfo {volume} {97}},\ \bibinfo {pages} {022302}
  (\bibinfo {year} {2018}{\natexlab{a}})}\BibitemShut {NoStop}%
\bibitem [{\citenamefont {Singh}\ \emph
  {et~al.}(2018{\natexlab{b}})\citenamefont {Singh}, \citenamefont {Singh},
  \citenamefont {Dorai},\ and\ \citenamefont {Arvind}}]{kd-pra-2018-3}%
  \BibitemOpen
  \bibfield  {author} {\bibinfo {author} {\bibfnamefont {A.}~\bibnamefont
  {Singh}}, \bibinfo {author} {\bibfnamefont {H.}~\bibnamefont {Singh}},
  \bibinfo {author} {\bibfnamefont {K.}~\bibnamefont {Dorai}}, \ and\ \bibinfo
  {author} {\bibnamefont {Arvind}},\ }\href {\doibase
  10.1103/PhysRevA.98.032301} {\bibfield  {journal} {\bibinfo  {journal} {Phys.
  Rev. A}\ }\textbf {\bibinfo {volume} {98}},\ \bibinfo {pages} {032301}
  (\bibinfo {year} {2018}{\natexlab{b}})}\BibitemShut {NoStop}%
\end{thebibliography}

%
\end{document}